\documentclass{article}

 \usepackage[eandd, final]{neurips_2026}
\usepackage[acronym]{glossaries}
\makeglossaries

\usepackage[utf8]{inputenc} % allow utf-8 input
\usepackage[T1]{fontenc}    % use 8-bit T1 fonts
\usepackage{hyperref}       % hyperlinks
\usepackage{url}            % simple URL typesetting
\usepackage{booktabs}       % professional-quality tables
\usepackage{amsfonts}       % blackboard math symbols
\usepackage{nicefrac}       % compact symbols for 1/2, etc.
\usepackage{microtype}      % microtypography
\usepackage{xcolor}         % colors
\usepackage{booktabs}
\usepackage{tabularx}
\usepackage{array}
\usepackage{pgf-pie}
\usepackage{listings}
\usepackage[most]{tcolorbox}
\usepackage{tcolorbox}
\usepackage{algorithm}
\usepackage{algpseudocode}
\usepackage{algorithmicx}
\usepackage{multirow}
\usepackage{xcolor}
\tcbuselibrary{listings, skins, breakable}
\usepackage{array}
\usepackage{pifont}
\usepackage{xspace}
\usepackage{listings}
\usepackage{caption}
\usepackage{wrapfig}
\tcbuselibrary{breakable}
\newacronym{llm}{LLM}{Large Language Model}
\newacronym{lm}{LM}{Language Model}
\newacronym{pro}{PRO}{Program-level Real-world Optimization}
\newacronym{pr}{PR}{Pull Request}
\newacronym{rciw}{RCIW}{Relative Confidence Interval Width}
\newacronym{rcihw}{RCIHW}{Relative Confidence Interval Half-Width}
\newacronym{cv}{CV}{Coefficient of Variation}
\newacronym{gc}{GC}{Garbage Collection}
\newacronym{auc}{AUC}{Area Under the Curve}
\newacronym{iid}{i.i.d.}{independent and identically distributed}
\newacronym{jvm}{JVM}{Java Virtual Machine}
\newacronym{jit}{JIT}{Just-In-Time Compilation}
\newacronym{ci}{CI}{Confidence Interval}
\newacronym{iqr}{IQR}{Interquartile Range}
\newacronym{twmu}{TWMU}{Time-Weighted Memory Usage}
\newacronym{pmu}{PMU}{Peak Memory Usage}
\newacronym{snr}{SNR}{Signal-to-Noise Ratio}
\newacronym{if}{IF}{Improvement Factor}
\newacronym{hlgf}{HLGF}{Human-LLM Gap Factor}
\newacronym{mad}{MAD}{Median Absolute Deviation}

\lstdefinestyle{arxivlisting}{
    basicstyle=\ttfamily\small,
    breaklines=true,
    breakatwhitespace=false,
    columns=fullflexible,
    keepspaces=true,
    showstringspaces=false,
    frame=none,
    numbers=none
}

\newtcblisting{pythonbox}[1]{
    listing engine=listings,
    listing options={style=arxivlisting, language=Python},
    colback=white,
    colframe=blue!60!black,
    fonttitle=\bfseries,
    title=#1,
    listing only,
    breakable,
    before skip=4pt,
    after skip=4pt
}

\newtcblisting{jsonbox}[1]{
    listing engine=listings,
    listing options={style=arxivlisting},
    colback=white,
    colframe=orange!60!black,
    fonttitle=\bfseries,
    title=#1,
    listing only,
    breakable,
    before skip=4pt,
    after skip=4pt
}

\newtcblisting{diffbox}[1]{
    listing engine=listings,
    listing options={style=arxivlisting},
    colback=teal!3,
    colframe=teal!70!black,
    fonttitle=\bfseries,
    title=#1,
    breakable,
    listing only,
    before skip=4pt,
    after skip=4pt
}

\newcommand{\bench}{SWE-Opt\xspace} 
\newcommand{\cmark}{\textcolor{green!50!black}{\ding{51}}}
\newcommand{\xmark}{\textcolor{red!60!black}{\ding{55}}}

\usepackage{xcolor}

\usepackage{xcolor}
\usepackage{colortbl}
\definecolor{refrow}{RGB}{255,242,204}

\title{\textbf{SWE}-Opt: A Benchmark for Evaluating LLMs on Performance-Oriented Repository-Level Optimization}

\author{%
  Ezgi Sarıkayak \\
  Siemens Research \\
  Technical University of Munich \\
  Munich, Germany \\
  \texttt{ezgi.sarikayak@siemens.com} \\
  \And
  Wenchao Gu\thanks{Corresponding author} \\
  Technical University of Munich \\
  Heilbronn, Germany \\
  \texttt{wenchao.gu@tum.de} \\
  \And
  Hesham Ghonim \\
  Technical University of Munich \\
  Munich, Germany \\
  \texttt{hesham.ghonim@tum.de} \\
  \And
  Chunyang Chen \\
  Technical University of Munich \\
  Heilbronn, Germany \\
  \texttt{chun-yang.chen@tum.de} \\
}

\begin{document}

\maketitle

\begin{abstract} 
Software performance optimization is a notoriously complex and manual task. Despite the growing use of \glspl{llm} for code refinement, we still lack benchmarks that capture how optimization actually happens in real-world codebases. Existing frameworks often oversimplify the problem by focusing on isolated functions or a single performance metric, missing the critical trade-offs between execution time and memory footprint, the inherent noise of the measurement environment, and the variability introduced by different input data and execution conditions. We address this by introducing \bench, a repository-level benchmark derived from 102 expert-written optimizations from open-source projects. Unlike previous benchmarks, \bench pairs each task with parameterized tests to evaluate runtime, peak memory, and \gls{twmu} across varying input data and execution conditions under 
noise-aware measurement conditions. Our evaluation shows that current \glspl{llm} struggle significantly: runtime gains are sparse and inconsistent, and memory optimizations are nearly non-existent. An agentic setting with iterative repository exploration improves patch correctness substantially but does not translate into performance gains.
This stands in sharp contrast to expert implementations, which achieve an aggregate speedup of $15.5\times$ and peak memory reduction of $171.3\times$ over benchmark tasks. Expert-written improvements are observed in $93.1\%$ of tasks for runtime and $70.6\%$ for peak memory. Our findings expose a substantial gap between current \gls{llm} capabilities and the demands of expert-level engineering.
\end{abstract}

\section{Introduction}
\label{introduction}
In recent years, \glspl{llm} have fundamentally reshaped software engineering, demonstrating remarkable proficiency in domains such as code generation~\cite{10.1145/3747588} and automated bug fixing~\cite{10.1016/j.csi.2024.103951}. As these models increasingly automate large-scale software production, the evaluative focus is pivoting from basic functional correctness toward long-term sustainability and execution efficiency. In this AI-driven development era, sub-optimal code can rapidly aggregate into "performance debt," significantly inflating computational overhead~\cite{computation_overhead}. Consequently, this technological shift has catalyzed interest in LLM-based code performance optimization—the critical task of refining implementations to enhance runtime and memory efficiency without altering functional semantics.

Despite its importance, performance optimization remains notoriously challenging~\cite{performance_bugs}. Unlike binary correctness, which provides deterministic pass/fail signals, optimization is inherently multidimensional and context-sensitive. A single optimization may involve conflicting trade-offs between runtime and memory usage, and its efficacy often fluctuates across different execution environments~\cite{10.1007/s10664-023-10391-y}. Moreover, performance is inextricably linked to workload characteristics; measurements are dynamic, context-dependent observations rather than static attributes~\cite{10.1145/3358960.3379130}. Given the increasing deployment of LLMs for such tasks, a rigorous benchmark is essential to evaluate their ability to navigate the complexities of real-world software performance.

Current evaluation paradigms have laid a foundational groundwork but face scalability and precision issues when transitioning from function-level to repository-level analysis. Early benchmarks like EffiBench~\cite{effibench} and COFFE~\cite{COFFE} assess individual functions in isolation, often neglecting the complex dependencies inherent in real-world repositories. While recent repository-level benchmarks such as SWE-Perf~\cite{sweperf}, SWE-FFICIENCY~\cite{SWE-fficiency}, and GSO~\cite{gso} operate at scale, they often lack sufficient evaluative rigor. Common pitfalls include using unit tests as a crude proxy for performance~\cite{sweperf} or aggregating heterogeneous execution variants into coarse-grained scores~\cite{gso}, which can obscure input-specific performance fluctuations. Such approaches offer a restricted view of how optimizations generalize across varying workloads (e.g., different input sizes, data properties, functional options)~\cite{muhlbauer_configurable} and typically lack the statistical rigor needed to distinguish genuine performance gains from environmental noise.

These limitations underscore a fundamental gap: existing benchmarks focus on task success under fixed conditions, failing to capture the input-dependent, statistically grounded, and multi-metric nature of modern performance optimization. This deficit motivates our investigation into whether LLMs can truly achieve robust repository-level optimization when subjected to rigorous, multi-workload scrutiny.

To address this, we present \bench \footnote{\bench is publicly available at \url{https://github.com/SWE-Opt/SWE-Opt}.}, a comprehensive repository-level benchmark comprising 102 expert-written optimization cases curated from real-world open-source projects. \bench is designed to shift the evaluation paradigm toward robust, workload-aware analysis. Unlike prior benchmarks, \bench evaluates tasks across diverse workloads via parameterized performance tests that reflect realistic execution scenarios. By conducting systematic parameter sweeps over input scales, data characteristics and function properties, our framework exposes performance behaviors that remain invisible to single-input benchmarks. To ensure reliability, we introduce an adaptive measurement mechanism: rather than relying on fixed-count profiling, \bench dynamically determines the necessary iterations based on the statistical convergence of the 90\% \gls{rciw}. Furthermore, we employ a multi-metric framework—incorporating peak memory and \gls{twmu} alongside runtime—to uncover critical time--memory trade-offs, providing a nuanced perspective on LLM-generated optimizations in resource-constrained contexts.

Through extensive experiments on state-of-the-art LLMs, we provide the first systematic study of repository-level code
optimization that jointly evaluates runtime and memory performance across varying workloads under noise-aware conditions. Our findings reveal that while modern models excel at local logic refinements, they struggle significantly to deliver measurable efficiency gains; their runtime improvements remain negligible, and memory optimizations are nearly non-existent. This stands in sharp contrast to expert implementations within our
benchmark, which achieve $15.48\times$ runtime speedup, $171.31\times$
peak memory reduction, and $619.22\times$ \gls{twmu} reduction, with
93.1\% of tasks showing reproducible runtime improvements, 70.6\%
showing reliable peak memory gains, and 55.9\% showing
\gls{twmu} improvements. Such a substantial performance gap suggests clear trajectories for future research in developing LLMs that can navigate the demands of expert-level engineering.

\section{\bench}
\label{probench_dataset}

\begin{table}[H]
\centering
\small
\caption{Comparison of repository-level performance optimization benchmarks. \textbf{Performance Scope}: how performance is evaluated per task. \textbf{Structured Parameter Sweep}: whether test cases are generated via structured sweeps over input dimensions. \textbf{Runtime} and \textbf{Memory}: performance metrics considered. \textbf{Adaptive Iterations}: whether measurement iterations are dynamically determined based on statistical convergence.}
\label{tab:benchmark_comparison}
\begin{tabularx}{\textwidth}{lXccccc}
\toprule
\textbf{Benchmark}
  & \textbf{\shortstack[l]{Performance\\Scope}}
  & \textbf{\shortstack{Structured\\Parameter Sweep}}
  & \textbf{Runtime}
  & \textbf{Memory}
  & \textbf{\shortstack{Adaptive\\Iterations}} \\
\midrule
GSO            & Multiple Scripts    & \xmark & \cmark & \xmark & \xmark \\
SWE-Perf       & Multiple Unit Tests & \xmark & \cmark & \xmark & \xmark \\
SWE-FFICIENCY  & Single Workload     & \xmark & \cmark & \xmark & \xmark \\
\midrule
\textbf{\bench} & Multiple Workloads & \cmark & \cmark & \cmark & \cmark \\
\bottomrule
\end{tabularx}
\end{table}

\subsection{Preliminaries}
\label{sec:Preliminaries}

\textbf{Formulation.} Repository-level performance optimization benchmarks task an agent with producing a patch to a given codebase that improves runtime efficiency while preserving functional correctness. SWE-Perf~\cite{sweperf} grounds tasks in performance-improving \glspl{pr}, pairing each instance with the full repository, a set of performance-related unit tests, and an expert-authored patch; evaluation reports patch applicability, functional correctness, and runtime improvement. 

Building on a similar construction, SWE-fficiency~\cite{SWE-fficiency} provides each agent with a complete codebase and a single fixed workload, requiring the agent to localize bottlenecks, edit the repository, and pass existing correctness tests; performance is quantified via the speedup ratio against a gold expert run on that workload configuration. In contrast, GSO~\cite{gso} constructs multiple performance tests per task from real commit histories, where each test is an experiment function that exercises the target code path across a range of input patterns and execution variants and is timed as a unit; performance is then evaluated via the OPT@K metric, measuring whether at least one of the top-K generated patches matches or exceeds the expert runtime.

\textbf{Limitations.} Current benchmarks are predominantly limited to runtime evaluation, entirely neglecting memory usage in their performance formulations. Furthermore, assessment under fixed or coarsely aggregated conditions significantly constrains the depth of analysis: SWE-Efficiency~\cite{SWE-fficiency} relies on a single workload per task, precluding any assessment of patch robustness across varying inputs; GSO~\cite{gso} conflates heterogeneous input variants into a single aggregate score, which obscures the origin of speedups and masks potential regressions at specific scales; and SWE-Perf~\cite{sweperf} employs pre-existing unit tests that were never designed to capture performance sensitivity. A systemic deficiency across these platforms is the absence of input sweeps and adaptive profiling, leaving results susceptible to measurement noise and preventing the identification of solutions that might fail under alternative configurations.

To address these shortcomings, \bench evaluates tasks across multiple structured workloads while jointly tracking runtime and memory consumption. It ensures measurement reliability through adaptive profiling, where the number of iterations is dynamically determined based on \gls{rciw} convergence. Moreover, \bench only accepts performance effects that exceed a noise-aware threshold, derived from both within-container variation and cross-container variation across repeated runs. The comprehensive advantages of \bench relative to existing benchmarks are summarized in Table~\ref{tab:benchmark_comparison}.

\subsection{Collection}
\label{sec:collection}

\textbf{Repository selection.} We focus on data-intensive Python libraries where performance optimizations are both frequent and measurable. To systematically identify optimization scenarios, we restrict our selection to repositories that maintain explicit performance-related labels on \glspl{pr}, enabling reliable extraction of developer-identified optimization tasks.

Based on these criteria, we select \texttt{pandas}, \texttt{scikit-learn}, and \texttt{xarray} as target repositories. These projects are widely used, actively maintained, and provide comprehensive unit test suites that enable reliable correctness validation. Further dataset statistics and repository-specific details are provided in Appendix~\ref{app:dataset-details}.

\textbf{PR filtering.}
We collect merged \glspl{pr} via the GitHub API and retain only those that (i) are explicitly labeled as performance-related, (ii) are merged into the main branch, (iii) involve modifications to Python source files, and (iv) include evidence of developer acknowledgment, such as entries in changelogs or \texttt{whatsnew} documentation. Requiring changelog or \texttt{whatsnew} entries acts as a proxy for significance, filtering out minor or incidental changes and retaining optimizations that are explicitly acknowledged by developers.

\textbf{Environment setup.}
For each \glspl{pr}, we reconstruct the baseline and optimized codebase states and execute them within isolated Docker environments. Each environment is
built using repository-specific dependencies corresponding to the target
commit, including development requirements to ensure compatibility with the test
suite.

The setup process requires careful inspection of repository
documentation and iterative refinement to ensure that the codebase can be
built and executed reliably. Using Docker ensures reproducibility and
consistent execution across workloads. Detailed Docker
configurations and reproducibility settings are provided in
Appendix~\ref{app:docker}.

\textbf{Unit Test Selection and Parameterized Performance Test Construction.}
For each \glspl{pr}, we identify unit tests associated with the target function and retain the full set to ensure comprehensive coverage of its functionality during evaluation.

Based on \glspl{pr} descriptions, code changes, and developer discussions, we construct parametrized performance test scenarios that capture the optimized operation. Each scenario defines an execution entry point and a parameterized space designed to trigger the performance behavior introduced by the \gls{pr}. Appendix~\ref{app:parameterized-perf}
details both the performance scenario construction process and the
formal parameter selection protocol, including parameter identification, workload generation, and implementation
examples.

\textbf{Validation.}
All instances are manually validated to remove inconsistent, non-reproducible, or unstable cases. We exclude \glspl{pr} with mixed modifications where the repository cannot be reliably built or tested in the reconstructed environment, ensuring reproducible tasks. 

\subsection{Features}
\label{sec:Features}

\textbf{Targeted scenario construction for optimization impact.}
Tasks are derived from performance-improving \glspl{pr}, grounding each instance in developer-identified optimization problems. Task descriptions are derived from \gls{pr} metadata, discussions, and code updates to define the optimization target and associated tests. They exclude hints and specify the target function and relevant input configurations.

\textbf{Parameterized performance tests.}
Each instance is evaluated under a parameterized setup defined by performance-relevant dimensions. We adopt a three-level abstraction: (i) \emph{parameters}, which define performance-relevant dimensions, (ii) \emph{parameter options}, which specify concrete values for each dimension, and (iii) \emph{execution configurations}, which specify how the performance-improving scenario is exercised with each combination in the parameter grid. This structure separates the definition of performance dimensions from their execution, enabling systematic and reproducible evaluation across diverse inputs. It allows analysis of how optimizations generalize and whether improvements hold across workloads or degrade under specific inputs.

\begin{figure}[t]
\centering
\hspace*{-0.08\textwidth}
\begin{minipage}[t]{0.58\textwidth}
\vspace{0pt}
\centering
\captionof{table}{Median characteristics of SWE-Opt task instances.}
\label{tab:task_median_statistics}
\small
\begin{tabular}{@{}lll@{}}
\toprule
\textbf{Category} & \textbf{Attribute} & \textbf{Median} \\
\midrule
Codebase & \# Python lines (non-test) & 197K \\
         & \# Python files (non-test) & 383 \\
\midrule
Gold Patch & \# Lines edited & 40 \\
           & \# Files edited & 3 \\
\midrule
\shortstack[l]{Correctness\\Tests} & \# Unit tests & 10 \\
\midrule
\shortstack[l]{Performance\\Tests} & \# Parameters & 2 \\
                                      & \# Execution configs. & 6 \\
\bottomrule
\end{tabular}
\end{minipage}%
\hspace{0.03\textwidth}%
\begin{minipage}[t]{0.32\textwidth}
\vspace{2.4em}
\centering
\begin{tikzpicture}
    \pie[
        text=legend,
        radius=1.30,
        sum=auto,
        color={teal!30, green!25, red!25}
    ]{
        79/\texttt{pandas},
        11/\texttt{scikit-learn},
        12/\texttt{xarray}
    }
\end{tikzpicture}
\captionof{figure}{Distribution of SWE-Opt instances across repositories.}
\label{fig:dataset-pie}
\end{minipage}
\end{figure}

\textbf{Dataset Distribution.} The final dataset consists of 102 validated benchmark instances collected 
across three repositories. Figure~\ref{fig:dataset-pie} shows the distribution 
across \texttt{pandas}, \texttt{scikit-learn}, and \texttt{xarray}, with 
\texttt{pandas} comprising the majority of instances. This reflects the 
availability of performance-labeled \glspl{pr} in open-source repositories 
rather than a deliberate selection bias. Table~\ref{tab:task_median_statistics} summarizes the median characteristics of each benchmark instance, including the size of the codebase, the size of the gold patch, and the structure of both correctness and parameterized performance tests.

To validate that each instance represents a genuine optimization opportunity, we evaluate gold solutions against their baselines using relative performance improvements in runtime, memory, and \gls{twmu}. We report only performance effects that pass \gls{snr} threshold, ensuring that observed differences are statistically distinguishable from measurement noise. Gold solutions achieve average runtime improvement of $15.48\times$ (93.1\% of tasks), a peak memory improvement of $171.31\times$ (70.6\% of tasks), and a \gls{twmu} improvement of $619.22\times$ (55.9\% of tasks).

\subsection{Reliable Performance Measurement Framework}
\label{sec:measurement-model}
Reliable performance measurement requires separating genuine optimization effects from pervasive environmental noise. We implement a hierarchical measurement framework to ensure that observed improvements in \bench are statistically significant and reproducible. For a more detailed discussion of our measurement methodology, see Appendix~\ref{app:measurement}.

\textbf{Hierarchical Noise Modeling.}
Performance measurements in containerized environments exhibit two independent sources of variance~\cite{kalibera_rigorous_reasonable}: \textit{within-container noise}, arising from scheduler jitter and interpreter behavior, and \textit{across-container noise}, arising from CPU state and memory layout variations. \bench addresses this by executing workloads across multiple fresh container instances, each performing adaptive measurement iterations inside.

\begin{figure}[t]
    \centering
    \includegraphics[width=0.80\linewidth]{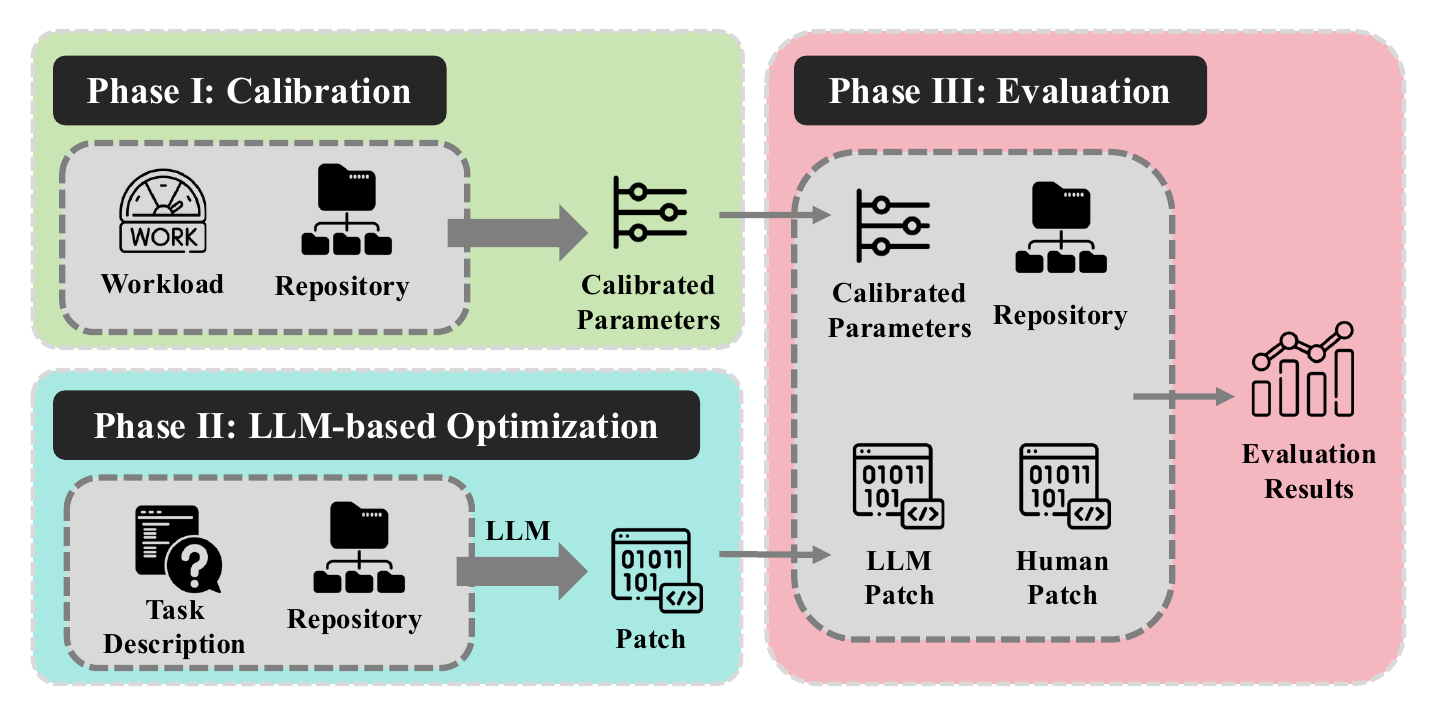}
    \caption{\bench performance measurement framework.}
    \label{fig:overview}
\end{figure}

\textbf{The Measurement Pipeline.}
As illustrated in Figure~\ref{fig:overview}, our pipeline consists of four integrated stages:
\begin{itemize}
\item \textbf{Calibration:} Performed once per workload to ensure comparability. For runtime, it determines the number of function \textit{invocations} per iteration to overcome timer resolution. For memory, it sets the \textit{sampling interval} required to capture allocation behavior with sufficient temporal resolution.
\item \textbf{Warmup:} To mitigate cold-start effects (e.g., cache initialization, lazy imports), we use a sliding-window convergence test based on the \gls{cv} of execution times. Measurement begins only after this criterion is met or a predefined limit is reached.
\item \textbf{Adaptive Sampling:} Unlike fixed-iteration benchmarks, we collect iterations until the \gls{rciw} for all metrics falls below a threshold ($\text{RCIW}_{0.9} < \epsilon$). This ensures that the uncertainty of the mean estimate is statistically bounded.
\item \textbf{Validation:} We apply two criteria—\textit{stability} (via RCIW) and \textit{detectability} (via \gls{snr} with a threshold of 2). Only measurements satisfying both are admitted for final reporting.
\end{itemize}

\textbf{Multi-dimensional Efficiency Metrics.}
Beyond runtime, \bench tracks memory consumption through two lenses: Peak Memory Usage above baseline and \gls{twmu}. Let $m(t)$ denote the sampled memory at time $t$, $m_{\mathrm{base}}$ the baseline footprint, and $T$ the duration. We define \gls{twmu} as:
\begin{equation}
\bar{m}_{\mathrm{twmu}} =
\frac{1}{T}\int_0^T \max(0, m(t)-m_{\mathrm{base}}) dt
\label{eq:twmu}
\end{equation}

This quantity is approximated via trapezoidal integration. By normalizing for execution time, \gls{twmu} provides a stable estimate of sustained memory pressure, preventing transient spikes from misrepresenting overall resource efficiency.

\section{Evaluating on \bench }
\label{sec:measurement}

\subsection{Experimental Setup}

We evaluate \bench{} across two complementary paradigms: single-shot, where the model generates a patch directly from the task description and provided context, and agentic, where the model iteratively explores the repository and refines its solution through multiple steps.

\textbf{Single-shot setting.}
We evaluate two context provision strategies:
\emph{oracle-based context} and \emph{BM25-based retrieval}.
The oracle strategy provides ground-truth relevant files
extracted directly from the target patch, while BM25
approximates this context using lexical similarity over
the repository state. This setup follows the methodology
introduced in SWE-bench~\cite{swebench}.
In both settings, the model receives the task description
and the selected repository context in a single prompt
and generates one patch without iterative refinement. These results establish a baseline of optimization capabilities under constrained, single-pass conditions. They should be interpreted as a capability floor rather than an upper bound on performance achievable through iterative refinement and tool-assisted exploration.
The prompt templates used for both strategies are provided
in Appendix~\ref{app:prompt-template}. Implementation details
for BM25 are provided in Appendix~\ref{app:retrieval}.

\textbf{Agentic setting.}
In addition to the single-shot settings, we evaluate
an iterative agentic setting using OpenCode~\cite{opencode}.
Unlike single-shot generation, the agent receives only
the task description and can autonomously explore
the target repository, execute commands, and iteratively
refine its modifications.
For each task, OpenCode operates in an isolated Docker
container derived from a SHA-pinned baseline image
containing the repository and its dependencies.
Following SWE-bench Pro~\cite{swebenchpro}, we limit
each agent run to a maximum of 50 steps.
Built-in web search and web fetching are disabled.
Parametrized performance tests and
reference patches are not exposed to the agent.
Upon termination, the final repository modifications
are extracted as a single Git patch against the
baseline commit and evaluated using the same
correctness and performance pipeline as the
single-shot settings.

\textbf{Models.}
We evaluate six large language models from multiple
providers in the single-shot settings:
GPT-5.2~\cite{gpt52} (OpenAI),
Claude Sonnet 4.6~\cite{claudesonnet46} (Anthropic),
Kimi K2.5~\cite{kimi25} (Moonshot AI),
Gemini 3.1 Flash~\cite{gemini31} (Google),
GLM-5.1~\cite{glm5} (Z.ai), and
MiniMax M2.7~\cite{minimaxm27} (MiniMax).
For the agentic setting, we restrict the evaluation
to three models: GPT-5.2, Claude Sonnet 4.6,
and Gemini 3.1 Flash.

\textbf{Evaluation metrics.}
We evaluate model performance along three dimensions:
(i) patch generation success, and 
(ii) functional correctness, and
(iii) performance.

\emph{Patch success and correctness.} Patch success measures whether a
generated patch can be applied without error. Correctness is evaluated
using the associated test suite, where a patch is considered correct if
all relevant unit tests pass after application. Both are defined over a task
set $\mathcal{T}$ as
\[
  \text{Patch@\cmark} = \frac{1}{|\mathcal{T}|} \sum_{t \in \mathcal{T}}
  \mathbb{I}[\text{patch applied}(t)],
  \qquad
  \text{Test@\cmark} = \frac{1}{|\mathcal{T}|} \sum_{t \in \mathcal{T}}
  \mathbb{I}[\text{tests pass}(t)].
\]

\emph{Performance metrics.}
Performance is evaluated only on tasks for which the patch applies successfully
and passes all correctness tests. We report three metrics: runtime,
peak memory usage, and \gls{twmu}, capturing execution speed, peak memory footprint, and total memory pressure over time respectively. Each metric quantifies a resource cost, where a reduction in the optimized version relative to the baseline constitutes an performance improvement.

We assess model effectiveness using two complementary metrics: \emph{Pass Rate} and \emph{\gls{if}}. Pass rate measures how frequently a model produces a valid optimization that satisfies the statistical filtering criteria. In contrast, IF measures the magnitude of performance changes on qualifying tasks.

We express performance changes through the \gls{if}, defined identically for all three resource metrics. Only workloads passing the \gls{snr} criterion are retained. This ensures that reported effects are statistically distinguishable from measurement noise. For each \gls{pr}, workload-level \gls{if} values are aggregated using the harmonic mean~\cite{harmonic} to obtain a single \gls{pr}-level score. We then report the arithmetic mean of these scores across all qualifying \glspl{pr}:
\[
\text{IF}_{\text{LLM}} =
\frac{1}{|\mathcal{P}|}
\sum_{\text{PR} \in \mathcal{P}}
\mathcal{H}\Bigl(
  \bigl\{\,
    \tfrac{\text{baseline}_{i}}{\text{LLM}_{i}}
  \bigr\}_{i \in \mathcal{W}_{\text{PR}}}
\Bigr),
\]
where $\mathcal{W}_{\text{PR}}$ denotes the set of workloads associated with a given \gls{pr}, and $\mathcal{P}$ denotes the set of \glspl{pr} whose aggregated improvement satisfies both the stability criterion based on \gls{rciw} and the minimum \gls{snr} threshold. The operator $\mathcal{H}$ denotes the harmonic mean used to aggregate workload-level improvement factors into a single \gls{pr}-level value. A value of $\text{IF}_{\text{LLM}} > 1$ indicates an overall performance improvement across workloads, whereas $\text{IF}_{\text{LLM}} < 1$ indicates that regressions outweigh improvements. 

Pass rate and \gls{if} should be interpreted jointly. A model with a high Pass rate may achieve frequent but modest improvements. Conversely, a model with a lower Pass rate may deliver substantially larger gains on fewer tasks. Thus, Pass rate captures the frequency of successful optimizations, while \gls{if} captures their magnitude. Neither metric alone provides a complete assessment of model optimization effectiveness. Further details on SNR-based filtering and results aggregation are provided in Appendix~\ref{app:measurement-phases} and~\ref{app:validation_aggregation}.

\section{Results} 
\label{results}

\subsection{Single-Shot Evaluation}
\label{sec:single-shot-results}

\begin{figure}[htbp]
    \centering
    \includegraphics[width=\linewidth]{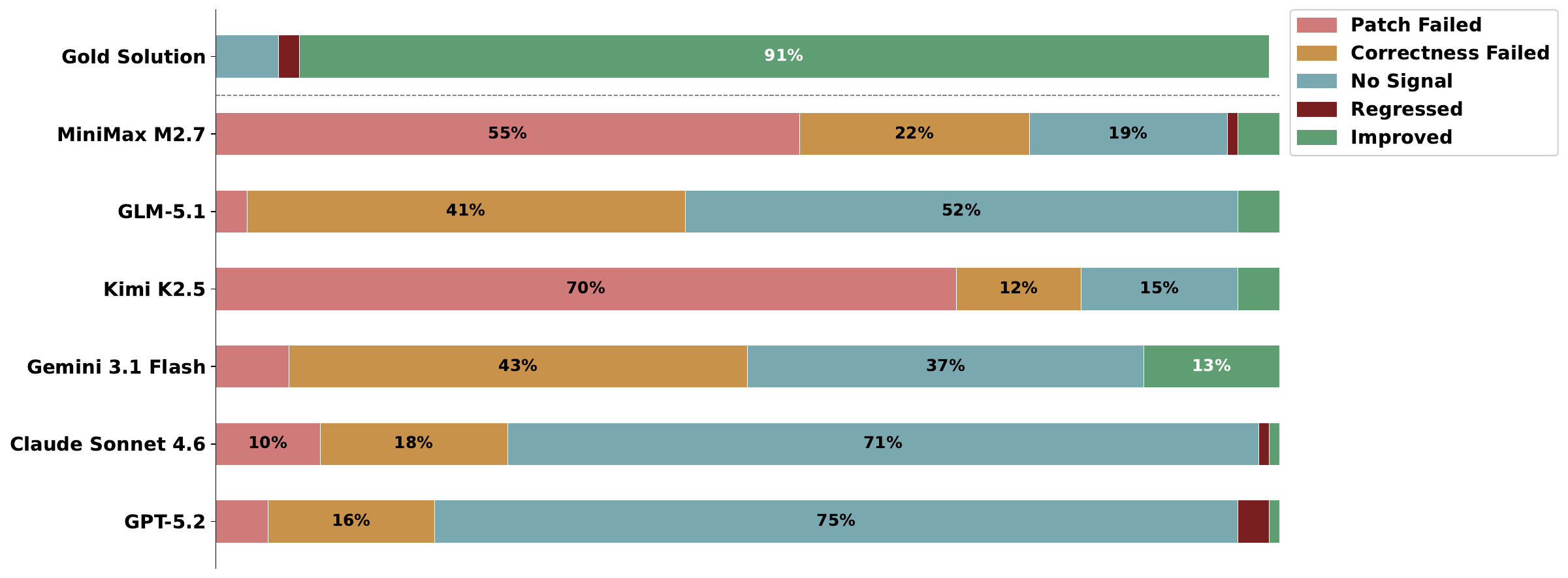}
    \caption{Task progression through the evaluation pipeline for each model
under Oracle retrieval. Patch application and correctness failures are
independent of performance evaluation. The remaining segments show
runtime outcomes. The Gold solution is included as reference.}
\label{fig:funnel}
\end{figure}

Producing reliable performance optimizations requires \glspl{llm} to satisfy three key conditions: generating patches that can be applied successfully, preserving functional correctness, and producing measurable performance improvements that survive \gls{snr} filtering. Figure~\ref{fig:funnel} shows the proportion of generated patches that satisfy each condition for different models, highlighting where optimization attempts fail. Table~\ref{tab:results-performance} further reports pass rate and \gls{if} under both Oracle and BM25 retrieval settings.

\textbf{\glspl{llm} can generate syntactically correct optimized code but struggle to maintain functional consistency.} Under Oracle retrieval, patch application rates range from 30.4\% to 97.1\% across models, while correctness rates are ranging from 18.6\% to 79.4\%. This gap shows that \glspl{llm} frequently produce syntactically applicable patches that nonetheless fail to preserve functional behavior. Under BM25 retrieval, both rates drop further, reflecting the sensitivity of patch quality to retrieval context. Full patch and correctness results under both retrieval settings are provided in Table~\ref{tab:results-correctness} in the Appendix.

\begin{table}[htbp]
\centering
\small
\caption{%
Pass rate and magnitude of detectable performance effects under Oracle and BM25 retrieval.
Each cell reports the percentage of all tasks passing the \gls{snr} filter and the corresponding \gls{if}.
\textbf{---} denotes that no \gls{pr} exhibited an \gls{snr} effect above the detectability threshold.
$^\ast$BM25 retrieval is not applicable to the Gold solution by design.}
\label{tab:results-performance}
\resizebox{\textwidth}{!}{%
\begin{tabular}{l l ccc ccc}
\toprule
& & \multicolumn{3}{c}{\textbf{Oracle}}
& \multicolumn{3}{c}{\textbf{BM25}} \\
\cmidrule(lr){3-5} \cmidrule(lr){6-8}
\textbf{Model} &
& \textbf{Runtime} & \textbf{Peak Mem.} & \textbf{TWMU}
& \textbf{Runtime} & \textbf{Peak Mem.} & \textbf{TWMU} \\
\midrule
\multirow{2}{*}{GPT-5.2}
  & Pass Rate & $3.9\%$  & ---          & $1.0\%$  & ---          & --- & --- \\
  & IF    & $0.69\times$ & ---      & $0.85\times$ & ---      & --- & --- \\[4pt]
\multirow{2}{*}{Claude Sonnet~4.6}
  & Pass Rate & $2.0\%$  & $1.0\%$      & $2.0\%$  & $12.7\%$     & $1.0\%$ & $2.0\%$ \\
  & IF    & $6.67\times$ & $16.61\times$ & $28.82\times$ & $2.28\times$ & $16.62\times$ & $3556.03\times$ \\[4pt]
\multirow{2}{*}{Gemini 3.1 Flash}
  & Pass Rate & $12.7\%$ & ---          & ---      & $13.7\%$     & --- & --- \\
  & IF    & $1.27\times$ & ---      & ---      & $1.19\times$ & --- & --- \\[4pt]
\multirow{2}{*}{Kimi K2.5}
  & Pass Rate & $3.9\%$  & ---          & ---      & $2.9\%$      & --- & --- \\
  & IF    & $1.18\times$ & ---      & ---      & $1.20\times$ & --- & --- \\[4pt]
\multirow{2}{*}{GLM-5.1}
  & Pass Rate & $3.9\%$  & ---          & ---      & $5.9\%$      & --- & --- \\
  & IF    & $1.16\times$ & ---      & ---      & $1.27\times$ & --- & --- \\[4pt]
\multirow{2}{*}{MiniMax M2.7}
  & Pass Rate & $4.9\%$  & ---          & $1.0\%$  & $2.0\%$      & --- & --- \\
  & IF    & $1.14\times$ & ---      & $1.11\times$ & $1.23\times$ & --- & --- \\
\midrule
\rowcolor{refrow}
\textbf{Gold solution} & Pass Rate & $93.1\%$ & $70.6\%$ & $55.9\%$ & $^\ast$ & $^\ast$ & $^\ast$ \\
\rowcolor{refrow}
 & IF & $15.48\times$ & $171.31\times$ & $619.22\times$ & $^\ast$ & $^\ast$ & $^\ast$ \\
\bottomrule
\end{tabular}%
}
\end{table}

\textbf{Even LLM-generated patches can pass correctness checks, most of them fail to produce signal beyond noise.} Even among tasks that pass both patch and correctness gates, the no-signal
segment dominates across all models in Figure~\ref{fig:funnel}. The
majority of patches that preserve correctness produce no performance change
that passes the \gls{snr} filter, meaning the observed differences cannot
be distinguished from measurement noise. This holds across all three
metrics and both retrieval settings. The Gold solution, in contrast, shows that 93.1\% of tasks admit measurable and reliable runtime improvements, confirming that the benchmark contains substantial optimizable structure that is not being captured by \glspl{llm}.

\textbf{Even LLM-generated patches can produce signal beyond noise, the gains are sparse and modest.}
Under Oracle retrieval, Gemini~3.1 Flash achieves the highest pass rate of
tasks with detectable runtime effects (12.7\%), while the corresponding
\gls{if} is $1.27\times$, indicating modest improvements in magnitude.
Claude Sonnet~4.6, in contrast, attains a lower runtime pass rate
(2.0\%) but achieves a larger \gls{if} of
$6.67\times$, suggesting that successful optimizations are less frequent
but occasionally more impactful. Under BM25, GPT-5.2 produces no reliable
runtime improvements. Claude Sonnet~4.6 remains the strongest model under
BM25 in terms of \gls{if}. In comparison, the Gold solution achieves a runtime \gls{if} of $15.48\times$ across most tasks, substantially exceeding all \gls{llm}-based improvements in both scale and consistency.

\textbf{Sometimes LLM-generated patches even introduce performance regressions.}
GPT-5.2 achieves a runtime \gls{if} of $0.69\times$ under Oracle retrieval,
indicating that the generated patches more often degrade performance than
improve it. A similar decline is observed for \gls{twmu} ($0.85\times$).
These results show that \gls{llm}-generated optimizations can introduce
systematic regressions rather than measurable performance gains.

\begin{figure}[htbp] 
    \centering
    \includegraphics[width=0.80\linewidth]{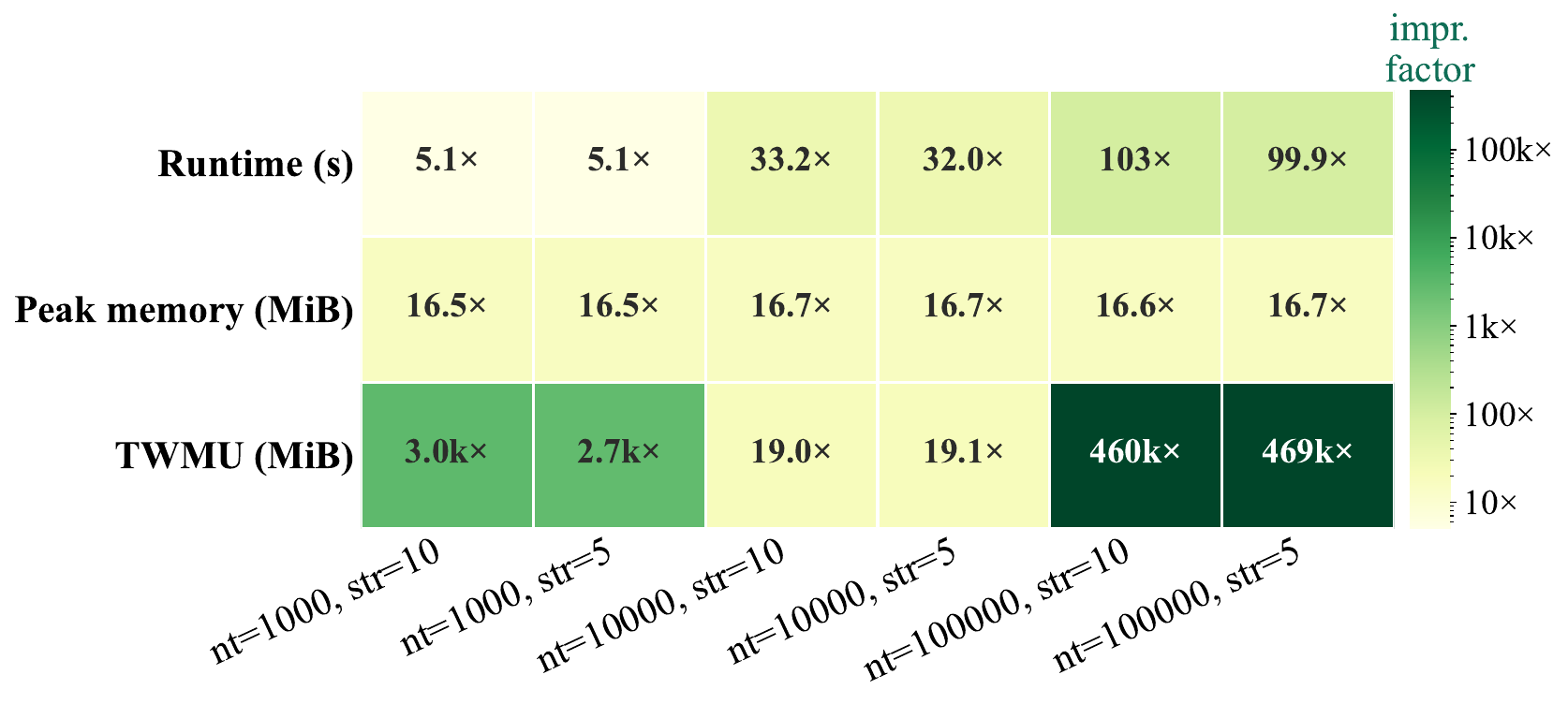}
    \caption{
Impact of input configuration on optimization effectiveness for PR~\#7578 under Oracle retrieval.
Per-workload \gls{if} varies with rolling size (\texttt{nt}) and stride
(\texttt{s}), highlighting input-dependent performance behavior.}
    \label{fig:llm_anaysis_claude_oracle}
\end{figure}

\textbf{Memory optimization remains largely beyond current \gls{llm} capabilities.}
Across models, peak memory and \gls{twmu} entries in
Table~\ref{tab:results-performance} are almost entirely absent, indicating
that consistent memory improvements are not achieved. The only notable
exception is Claude Sonnet~4.6 achieving a peak memory \gls{if} of $16.61\times$ and a \gls{twmu} \gls{if} of $28.82\times$ under Oracle retrieval. Under BM25 retrieval, the same model attains an even higher \gls{twmu} \gls{if} of $3556.03\times$, surpassing the Oracle result. However, these improvements are entirely driven by this single instance and do not generalize across tasks. This suggests that while \gls{llm}s can achieve strong memory optimizations in isolated cases, such behavior is not yet reliable or systematic. The Gold solution, however, consistently achieves large improvements across all metrics, including peak memory ($171.31\times$) and \gls{twmu} ($619.22\times$), indicating that these gains are broadly attainable when guided by expert-level reasoning rather than model-generated heuristics.

Figure~\ref{fig:llm_anaysis_claude_oracle} provides a detailed analysis of
PR\#7578, responsible for the large peak-memory and \gls{twmu} improvements,
by showing per-workload \gls{if} across varying input sizes for all three
metrics. Runtime improvement increases substantially with input size, reaching
around $100\times$ for large workloads. Peak-memory improvement remains
relatively stable at approximately $16\times$, while \gls{twmu} exhibits the
largest gains at scale, surpassing $460\times$ for the largest workloads. This
workload-dependent behavior highlights the importance of workload-relevant
profiling, as optimization effects may only become visible under sufficiently
large or stressed inputs. In contrast, Gold solutions achieve improvements
across most tasks rather than relying on a single extreme case, resulting in
broader and more consistent gains overall. Detailed results are provided in
Appendix~\ref{app:pr7578}.

% \textbf{Expert solutions establish a clear upper bound.}
% The Gold solution achieves a runtime \gls{if} of $15.48\times$,
% a peak-memory \gls{if} of $171.31\times$, and a \gls{twmu} \gls{if} of
% $619.22\times$, while producing reliable runtime improvements on 93.1\%
% of tasks. In contrast, \gls{llm}-generated patches achieve substantially
% lower pass rates and improvement magnitudes across all metrics. This
% large gap suggests that current \glspl{llm} still lack the deeper
% performance reasoning and structural understanding required to reliably
% identify and implement expert-level optimizations.

\subsection{Agentic Evaluation}
\label{sec:agentic-results}

We further investigate whether autonomous repository
exploration and iterative patch refinement improve
optimization outcomes compared to single-shot generation.
We evaluate three models using OpenCode and assess their
final patches using the same correctness and performance
pipeline as the single-shot settings.

\begin{table}[b]
\centering
\small\caption{Patch applicability, correctness tests, and performance effects under agentic execution.}
\label{tab:agentic-results}
\resizebox{\linewidth}{!}{%
\begin{tabular}{lcc*{3}{cc}}
\toprule
\multirow{2}{*}{\textbf{Model}}
& \multirow{2}{*}{\textbf{Patch@\cmark}}
& \multirow{2}{*}{\textbf{Test@\cmark}}
& \multicolumn{2}{c}{\textbf{Runtime}}
& \multicolumn{2}{c}{\textbf{Peak Mem.}}
& \multicolumn{2}{c}{\textbf{TWMU}} \\
\cmidrule(lr){4-5}
\cmidrule(lr){6-7}
\cmidrule(lr){8-9}
& & & Pass Rate & IF & Pass Rate & IF & Pass Rate & IF \\
\midrule

GPT-5.2
& 90.2\% & 85.3\%
& 14.7\% & $1.18\times$
& ---    & ---
& 1.0\%  & $1.11\times$ \\[3pt]

Claude Sonnet~4.6
& 96.1\% & 84.3\%
& 14.7\% & $25.29\times$
& 2.0\%  & $9.61\times$
& 2.9\%  & $2662.14\times$ \\[3pt]

Gemini 3.1 Flash
& 100.0\% & 94.1\%
& 17.6\% & $1.29\times$
& ---     & ---
& 1.0\%  & $1.11\times$ \\[3pt]

\midrule
\rowcolor{refrow}
\textbf{Gold solution}
& $^\ast$ & $^\ast$
& 93.1\% & $15.48\times$
& 70.6\% & $171.31\times$
& 55.9\% & $619.22\times$ \\
\bottomrule
\end{tabular}%
}
\end{table}

\textbf{Agentic execution and functional correctness.}
Table~\ref{tab:agentic-results} shows that patch application and
functional correctness rates improve substantially compared to
single-shot settings. Application rates range from 90.2\% 
to 100.0\% , and correctness rates reach up to
94.1\%, compared to a maximum of 79.4\% under single-shot setting.
Autonomous repository exploration and iterative refinement thus help
models produce more applicable and functionally consistent patches.

\textbf{Agentic execution and performance improvements.}
Despite these gains in correctness, the improvement does not carry over
to performance optimization. Runtime pass rates remain modest across all
models (14.7\%--17.6\%), and improvement magnitudes are largely
comparable to the single-shot Oracle setting. Claude Sonnet~4.6 is the
only exception, achieving a runtime \gls{if} of $25.29\times$ and
isolated memory gains, but these remain driven by individual cases
rather than consistent behavior across different tasks. The gap with the Gold
solution which achieves a 93.1\% runtime pass rate and a
$15.48\times$ \gls{if} highlights that iterative execution alone is
insufficient for reliable performance optimization, and that the
bottleneck lies in performance reasoning.
\section{Related Work}
\label{background}

\textbf{LLMs for Code Optimization.} Recent research has extensively 
explored \gls{llm}s for improving code efficiency through strategies 
such as performance-aware fine-tuning, iterative refinement, and 
search-based optimization~\citep{shypula2024learningperformanceimprovingcodeedits, 
10.1109/ICSE55347.2025.00021, 11052794}. Despite these advancements, 
achieving robust generalization and, in particular, reliable 
performance evaluation remains an open challenge~\cite{gong2025languagemodelscodeoptimization}. 
To assess optimization quality, existing evaluation frameworks have 
progressed across multiple levels of granularity, ranging from 
function-level benchmarks that abstract away system interactions 
(e.g., EffiBench, ENAMEL, ECCO, Mercury)~\cite{effibench, effibenchx, Enamel, ECCO, Mercury}, 
to program-level kernels that capture controlled computational 
patterns~\cite{COFFE}, and more recently to repository-level 
benchmarks such as GSO~\cite{gso}, SWE-Perf~\cite{sweperf}, and 
SWE-Efficiency~\cite{SWE-fficiency}, which incorporate real-world 
software systems. However, across all levels, evaluation is still 
predominantly based on execution time under fixed configurations, 
limiting the ability to capture realistic performance behavior.

\textbf{Performance Measurement and Multi-Metric Tradeoffs.} This 
limitation is fundamentally rooted in the nature of performance 
itself. Performance optimization is inherently multidimensional: 
improvements in one metric, such as runtime, do not necessarily 
translate to gains in others, such as memory usage or energy 
consumption, requiring careful navigation of tradeoffs in real-world 
systems~\citep{10.1007/s10664-023-10391-y, Kempen2024ItsNE, rui, 
10.1145/3788089, multiobjective}. Moreover, system behavior is tightly 
coupled with workload characteristics and execution context. 
Variations in input size, data distribution, and hardware conditions 
can shift bottlenecks and invalidate previously effective 
optimizations~\cite{muhlbauer_configurable, input_sens, perfgen}. 
Reliable performance measurement further introduces challenges due to 
non-determinism from hardware effects, OS scheduling, and runtime 
compilation, making naive evaluation statistically 
unsound~\cite{Mytkowicz_wrong_data, georges_rigorous, 
kalibera_rigorous_reasonable}. Even in controlled microbenchmark 
settings, optimization decisions must balance execution speed with 
result fidelity~\cite{laaber_dynamic_reconfigure}. Consequently, 
performance cannot be adequately characterized by a single metric or 
a static configuration; instead, it must be evaluated across multiple 
resource dimensions and workload variations to reflect true system 
behavior~\cite{10.1007/s10664-023-10391-y, muhlbauer_configurable, 
Mytkowicz_wrong_data, multiobjective}.

\section{Limitations}
\label{sec:limitations}
\bench\ focuses on Python-level optimization in data-intensive libraries where workloads can be varied through input parameters. The benchmark includes only repositories with explicitly performance-labeled \glspl{pr}, ensuring reproducibility but limiting coverage. Consequently, findings may not generalize to low-level systems software, I/O-bound applications, or non-Python environments. Extending \bench\ to multiple languages remains an important direction for future work.

\section{Conclusion}
\label{conclusion}
We presented \bench, a repository-level benchmark for evaluating \gls{llm} capabilities in real-world performance optimization. \bench\ comprises 102 expert-authored optimizations with parameterized performance tests covering runtime, peak memory, and \gls{twmu}. Our results reveal a substantial gap between current \glspl{llm} and expert developers: reliable runtime improvements are rare, memory optimizations are largely absent, and the majority of generated patches fail to produce any measurable gain. The primary bottleneck lies not in the magnitude of improvements when they occur, but in reliably identifying optimization opportunities in the first place. Expert solutions, by contrast, achieve strong and consistent improvements across tasks, establishing a challenging upper bound for future systems.

 \section*{Acknowledgements}

This work was supported by the German Federal Ministry of Research, Technology and Space (BMFTR) under ITEA4 project “23016 GreenCode” (16IS24070G)

\bibliographystyle{plain}
\bibliography{references}

\appendix

\section{Dataset Construction and Statistics}
\label{app:dataset-details}

This section provides implementation-level details of the dataset
construction process, complementing Section~\ref{probench_dataset}.
We focus on repository-specific configurations and the resulting
dataset composition.

\subsection{Pull Request Collection}
\label{app:pr-collection}

\glspl{pr} are retrieved via the GitHub API. We filter \glspl{pr} based on repository-specific performance-related labels and the presence of corresponding changelog updates, retaining only source-level changes to Python files. Merge commits are excluded to ensure that extracted changes correspond directly to the \gls{pr}.

While the collection procedure is consistent across repositories, configurations vary in label definitions, changelog locations. These repository-specific settings are summarized in Table~\ref{tab:pr-collection} with repository license. Table~\ref{tab:line_changes} further shows that benchmark instances span a wide range of change sizes, from small edits to substantial modifications, indicating diverse optimization complexity.

\begin{table}[h]
\centering
\caption{Repository-specific details for \gls{pr} collection.}
\label{tab:pr-collection}
\begin{tabular}{lcccc}
\toprule
\textbf{Criterion} & \textbf{pandas} & \textbf{scikit-learn} & \textbf{xarray} \\
\midrule
Performance label  & Performance & Performance & topic-performance \\
Changelog path     & doc/source/whatsnew & doc/whats\_new & doc/whats-new\\
License            & BSD 3-Clause & BSD 3-Clause & Apache 2.0 \\
\bottomrule
\end{tabular}
\end{table}

\begin{table}[h]
\centering
\caption{Summary statistics of code changes across benchmark instances.}
\label{tab:line_changes}
\begin{tabular}{lrrrr}
\toprule
\textbf{Metric} & \textbf{Min} & \textbf{Median} & \textbf{Mean} & \textbf{Max} \\
\midrule
Lines changed  &   3.0 &  39.5 &  77.7 & 1{,}424.0 \\
Files changed  &   2.0 &   3.0 &   3.6 &    14.0 \\
\bottomrule
\end{tabular}
\end{table}

\section{Construction and Parameterization of Performance Tests}
\label{app:parameterized-perf}

Parameterized performance tests are constructed from \gls{pr} descriptions, developer discussions, and associated code changes. These sources capture intended usage patterns and the conditions under which the optimization is effective.

\subsection{Workload Categories.}

Parameters are grouped into four categories:

\begin{itemize}
    \item \textbf{Input size:} Defines the scale of the input data (e.g., number of elements, rows, or features).
    
    \item \textbf{Data distribution:} Captures statistical and structural properties of the input, such as sparsity, duplication, or missingness, which influence execution behavior.
    
    \item \textbf{Data type:} Specifies the representation of the input data (e.g., numeric, categorical, or mixed), affecting internal processing and memory layout.
    
    \item \textbf{Function-level configuration:} Includes parameters and options that control the behavior of the target function independently of the input data.
\end{itemize}

\subsection{Workload Selection and Options.}

Parameters are selected to expose performance-relevant behavior while maintaining a tractable workload space. For each parameter, a set of \emph{parameter options} is defined, representing concrete values used to instantiate workloads.

The selection process is guided by the following criteria:

\begin{itemize}
    \item \textbf{Activation of optimized behavior:} Selecting values that trigger the optimized execution paths targeted by the \gls{pr}.
    
    \item \textbf{Performance sensitivity:} Including values that expose meaningful variation in runtime or memory behavior.
    
    \item \textbf{Coverage vs. complexity:} Limiting the number of parameter options to avoid combinatorial explosion while preserving coverage of critical behaviors.
    
    \item \textbf{Resource feasibility:} Ensuring that all workloads can be executed within runtime and memory constraints.
\end{itemize}

To ensure practical feasibility, each scenario is limited to at most 24 workloads. Workloads are generated by combining parameter options via a Cartesian product; therefore, the number of options per parameter is chosen such that the total number of resulting combinations remains within this bound. Table~\ref{tab:workload_space} summarizes the resulting workload configurations across all scenarios. 

\begin{table}[h]
\centering
\caption{Summary statistics of workload configuration space across 102 benchmark scenarios, reporting the distribution of parameter dimensions and workload configurations per scenario.}
\label{tab:workload_space}
\begin{tabular}{lrrrr}
\toprule
\textbf{} & \textbf{Min} & \textbf{Median} & \textbf{Mean} & \textbf{Max} \\
\midrule
Parameter dimensions   & 1 & 2.0 & 1.9 & 4 \\
Workload configurations & 2 & 6.0 & 6.9 & 24 \\
\bottomrule
\end{tabular}
\end{table}

\subsection{Parametrized Performance Test Implementation.}
Each performance test consists of three components:

\begin{itemize}
    \item \texttt{setup()}: constructs inputs and configuration from the workload parameter set,
    \item \texttt{warmup()} (optional): stabilizes cache-sensitive behavior,
    \item \texttt{run()}: invokes only the target API.
\end{itemize}

Isolating the target call in \textbf{run()} ensures that only the operation of interest is profiled.

Each test is accompanied by a \textbf{workloads} field that defines the parameter space. Each parameter entry specifies 
(i) a name, 
(ii) a short identifier, 
(iii) a set of admissible values, and 
(iv) a brief description of its role.

Inputs are generated using a fixed random seed to ensure reproducibility. The following examples illustrate one parameterized performance test from each of the three target repositories, including their corresponding workloads.

\paragraph{pandas --- PR\#50778}

PR\#50778 optimizes \texttt{Series.replace} when using a large
dictionary for substitution. The test varies the series length
(\texttt{N}) and the number of values to replace (\texttt{R}) to
expose how performance scales with dictionary size and series length.

\begin{pythonbox}{Performance Scenario --- pandas PR\#50778}
import numpy as np
import pandas as pd
from swe_opt.scenarios.scenario_base import PerfScenario

class PR50778Scenario(PerfScenario):
    """
    PR #50778 - Benchmark Series.replace when using a large dict
    for to_replace.

    Parameters:
    - N: Series length
    - R: Number of values to replace
    """
    def setup(self):
        N = int(self.params["N"])
        R = int(self.params["R"])

        arr = self.rng.standard_normal(N)
        arr1 = arr.copy()
        self.rng.shuffle(arr1)

        ser = pd.Series(arr)
        to_replace = self.rng.choice(arr, R)
        values    = self.rng.choice(arr1, R)

        self.args = {
            "series": ser,
            "replace_dict": dict(zip(to_replace, values))
        }

    def run(self):
        return self.args["series"].replace(self.args["replace_dict"])
\end{pythonbox}
\begin{jsonbox}{Parameter Configuration --- pandas PR\#50778}
"params": [
  {
    "name": "Series Length",
    "short_indicator": "N",
    "values": ["100000", "500000", "1000000"],
    "explanation": "Number of elements in the Series"
  },
  {
    "name": "Number to Replace",
    "short_indicator": "R",
    "values": ["1000", "5000"],
    "explanation": "Number of elements to replace"
  }
]
\end{jsonbox}

\paragraph{scikit-learn --- PR\#21808}

PR\#21808 improves \texttt{LogisticRegression} with the
\texttt{lbfgs} solver for multiclass classification. The scenario
varies the number of samples (\texttt{n}), features (\texttt{p}),
and classes (\texttt{c}), as the optimization benefit increases
with the number of classes.

\begin{pythonbox}{Performance Scenario --- scikit-learn PR\#21808}
from sklearn.linear_model import LogisticRegression
from swe_opt.scenarios.scenario_base import PerfScenario

class PR21808Scenario(PerfScenario):
    """
    PR #21808 - Benchmark LogisticRegression.fit with lbfgs solver
    for multiclass classification.

    Parameters:
    - n: Number of samples
    - p: Number of features
    - c: Number of classes
    """
    def setup(self):
        n = int(self.params["n"])
        p = int(self.params["p"])
        c = int(self.params["c"])

        X = self.rng.standard_normal((n, p))
        y = self.rng.integers(0, c, size=n)

        self.args = {
            "model": LogisticRegression(
                solver="lbfgs",
                max_iter=100,
                multi_class="multinomial"
            ),
            "X": X,
            "y": y
        }

    def run(self):
        return self.args["model"].fit(self.args["X"], self.args["y"])
\end{pythonbox}
\begin{jsonbox}{Parameter Configuration --- scikit-learn PR\#21808}
"params": [
  {
    "name": "n_samples",
    "short_indicator": "n",
    "values": ["50000", "100000"],
    "explanation": "Number of samples"
  },
  {
    "name": "n_features",
    "short_indicator": "p",
    "values": ["50", "100"],
    "explanation": "Number of features"
  },
  {
    "name": "n_classes",
    "short_indicator": "c",
    "values": ["5", "10", "20"],
    "explanation": "Number of classes (benefit increases with more classes)"
  }
]
\end{jsonbox}

\paragraph{xarray --- PR\#7578}
\label{par:xarray_7578}
PR\#7578 optimizes \texttt{DatasetRolling.construct} with stride on
datasets without indexes. The workloads varies the size of the
rolling dimension (\texttt{nt}) and the stride (\texttt{str}) to
capture how performance scales with both rolling window size and
step size.

\begin{pythonbox}{Performance Scenario --- xarray PR\#7578}
import numpy as np
import xarray as xr
from swe_opt.scenarios.scenario_base import PerfScenario

class PR7578Scenario(PerfScenario):
    """
    PR #7578 - Benchmark DatasetRolling.construct with stride on
    Datasets without indexes.

    Parameters:
    - nt:  Size of rolling dimension
    - str: Stride for construct
    """
    def setup(self):
        nt     = int(self.params["nt"])
        stride = int(self.params["str"])

        data_vars = {
            "temperature": (["time", "space"],
                self.rng.standard_normal((nt, 100))),
            "pressure":    (["time", "space"],
                self.rng.standard_normal((nt, 100))),
        }
        coords = {
            "time":         np.arange(nt),
            "space":        np.arange(100),
            "time_label":   ("time", [f"t_{i}" for i in range(nt)]),
            "space_weight": ("space", self.rng.random(100)),
        }

        ds = xr.Dataset(data_vars=data_vars, coords=coords)
        self.args = {
            "func":   ds.rolling(time=10),
            "stride": stride
        }

    def run(self):
        return self.args["func"].construct(
            "window_dim", stride=self.args["stride"]
        )
\end{pythonbox}
\begin{jsonbox}{Parameter Configuration --- xarray PR\#7578}
"params": [
  {
    "name": "n_time",
    "short_indicator": "nt",
    "values": ["1000", "10000", "100000"],
    "explanation": "Size of rolling dimension"
  },
  {
    "name": "stride",
    "short_indicator": "str",
    "values": ["5", "10"],
    "explanation": "Stride for construct"
  }
]
\end{jsonbox}

\section{Retrieval Configuration}
\label{app:retrival_details}

The two context provision strategies differ in how relevant code is identified and supplied to the model.

\textbf{Oracle Context.}
Oracle context is extracted offline during dataset construction. For each sample, the ground-truth files modified by the developer patch are identified and stored directly in the dataset. At inference time, no retrieval is performed, the relevant context is loaded directly from the precomputed dataset record.

\textbf{BM25 Retrieval.}
\label{app:retrieval}
BM25 context is retrieved at runtime from the repository checked out at the baseline commit. The repository is indexed using Pyserini Lucene backend~\cite{pyserini}, and retrieval is performed using a query constructed from the task description. All experiments use a fixed BM25 configuration with $k_1 = 1.0$ and $b = 0.3$, following prior work on code retrieval~\cite{bm25_param_study}.

\section{Prompt Construction: Template and Examples}
\label{app:prompt-template}

We describe the prompt construction used in the single-shot and agentic
settings. The two settings differ fundamentally in how context is
provided and how output is constrained.

\subsection{Single-Shot Prompt}

Each single-shot prompt consists of three components: (i) a task
description specifying the optimization objective and constraints,
(ii) a code context providing the relevant source code, and (iii) a
structured output specification that constrains the model to produce
search-and-replace code edits.

\begin{verbatim}
<task description>

[start of src/module/example.py]
### Code context
```python
# complete file source
```
[end of src/module/example.py]

<response instruction>
\end{verbatim}

\textbf{Task Description.}
The task description defines a fixed optimization objective, followed by explicit behavioral constraints and a task-specific goal derived from \gls{pr} metadata. It captures both the target function and the input conditions under which performance issues arise (e.g., large inputs, sparsity, or missing values), reflecting the developer’s intended optimization.

\begin{verbatim}
You are an expert Python performance engineer. Your objective is to optimise
<entry_point> for speed and memory efficiency without altering its behaviour,
public APIs, or test expectations.

CONSTRAINTS:
- Preserve exact semantics and outputs.
- Preserve all public APIs and return types.
- Do not modify tests or test expectations.

OPTIMIZATION REQUEST:
<task description>
Optimization function: <entry_point>
\end{verbatim}

\textbf{Response Instruction.} The response instruction defines a structured output format that constrains the model to generate deterministic search-and-replace code edits.

\begin{verbatim}
OUTPUT FORMAT — SEARCH/REPLACE BLOCKS:

For each change emit exactly one block:

path/to/file.py
[SEARCH]
<code to find>
[/SEARCH]
[REPLACE]
<replacement code>
[/REPLACE]

RULES:
- Output ONLY the blocks. No prose, no explanations, no markdown fences.
- Multiple blocks are allowed — one per logical change, in any order.
- SEARCH must be copied VERBATIM from the file:
    * exact line breaks — never collapse a multiline signature to one line
    * exact indentation — do not add or remove leading whitespace
    * exact quotes — do not change ' to " or vice versa
    * exact spacing — do not add or remove spaces around = or ()
- SEARCH must be the SMALLEST snippet that uniquely identifies the location.
  Never include whole functions or docstrings unless the whole thing changes.
- REPLACE may be empty to delete the matched code.
- Multiple blocks on the same file are applied top-to-bottom.

ANCHORING GUIDANCE:
- REPLACE: SEARCH = the lines to change, REPLACE = the new lines.
- ADD:     SEARCH = nearest unique surrounding line(s), REPLACE = those same
           lines plus the new code inserted before or after.
- DELETE:  SEARCH = the lines to remove, REPLACE = (empty).
- If a function signature spans multiple lines in the file, your SEARCH must
  span those exact same lines — do not rewrite it as a single line.

EXAMPLES:

# 1. Replace a function, method, or class body
mypackage/module.py
[SEARCH]
    def compute(self, x):
        return x * 2
[/SEARCH]
[REPLACE]
    def compute(self, x: int) -> int:
        return x * 3
[/REPLACE]

# 2. Replace a multiline signature
mypackage/module.py
[SEARCH]
    def process(
        self,
        value: int,
    ) -> int:
[/SEARCH]
[REPLACE]
    def process(
        self,
        value: int,
        scale: float = 1.0,
    ) -> int:
[/REPLACE]

# 3. Add a function, method, or class after an existing one
mypackage/module.py
[SEARCH]
    def existing_method(self):
        pass
[/SEARCH]
[REPLACE]
    def existing_method(self):
        pass

    def new_method(self):
        return True
[/REPLACE]

# 4. Delete a function, method, or class
mypackage/module.py
[SEARCH]
    def deprecated_method(self):
        do_old_thing()
        return result
[/SEARCH]
[REPLACE]
[/REPLACE]
\end{verbatim}

\subsection{Agentic Prompt}

The agentic prompt differs from the single-shot prompt. The agent receives only the task
description, entry point, and benchmark-wide optimization
objectives, and must locate the relevant implementation by
exploring the repository autonomously.

\begin{verbatim}
You are an expert software performance engineer.

OBJECTIVE:
Optimize the requested operation according to all three benchmark metrics:
- Runtime: reduce execution time.
- Peak memory: reduce the maximum memory used during execution.
- Time-weighted mean memory: reduce memory usage over the full execution period.

Treat all three metrics as optimization objectives. Preserve externally
observable behavior, and do not improve one metric through a substantial
regression in another.

BENCHMARK TASK:
<task description>

ENTRY POINT:
<entry_point>

CONSTRAINTS:
- Preserve exact observable behavior and outputs.
- Preserve public APIs, signatures, and return types.
- Preserve existing test expectations.
- Avoid changes unrelated to the requested optimization.

WORKING ENVIRONMENT:
- The current working directory is the repository root.
- The repository is checked out at the required baseline commit.
- Project dependencies and test tools are already installed.
- Do not install or update dependencies.
- Do not modify tests, benchmarks, fixtures, or expected outputs.
- Leave only source-code changes required for the optimization.

BUDGET:
You have a maximum of 50 agent turns.
\end{verbatim}

\section{Measurement Pipeline Details} %todo here
\label{app:measurement}

This appendix provides the full algorithmic specification and parameter settings of the measurement pipeline described in Section~\ref{sec:measurement}. Each subsection corresponds to a stage of the pipeline and presents the associated implementation details.

% ── Configuration table ──────────────────────────────────
\subsection{Configuration Parameters}
\label{app:params}

This subsection summarizes the configuration parameters of the measurement pipeline, grouped by the warmup, calibration, and measurement stages. For each parameter, we report its value and purpose. Table~\ref{tab:measure-config} lists all parameters using notation consistent with Algorithms~\ref{alg:calibration}--\ref{alg:measurement}, ensuring alignment between the specification and implementation.

The parameter values are chosen to balance measurement stability and computational cost. We follow established principles of rigorous benchmarking, including warmup execution, repeated measurements, and variance-aware stopping criteria~\cite{kalibera_rigorous_reasonable,georges_rigorous,laaber_dynamic_reconfigure}. Parameters not prescribed by prior work are set empirically based on pilot experiments and kept fixed across all scenarios to ensure consistency and comparability.

For runtime measurement, parameters control the number of invocations per iteration and the minimum sample duration, ensuring that measurements are not dominated by timer resolution or short-term system noise. For memory measurement, parameters define the sampling configuration used during tracing, with values chosen to provide sufficient temporal resolution while limiting instrumentation overhead.

Finally, parameters related to convergence and validation are configured to ensure that collected measurements are both stable and detectable performance effects. The rationale behind these criteria is discussed separately in Appendix~\ref{app:measurement-phases} and Appendix~\ref{app:validation_aggregation}.

\begin{table}[!tbp] 
\centering
\caption{Measurement pipeline configuration and parameters used in Algorithms~\ref{alg:calibration}--\ref{alg:measurement}.}
\label{tab:measure-config}
\footnotesize
\setlength{\tabcolsep}{4pt}
\renewcommand{\arraystretch}{1.05}
\begin{tabularx}{\linewidth}{p{0.20\linewidth} p{0.18\linewidth} p{0.10\linewidth} X}
\toprule
\textbf{Category} & \textbf{Parameter} & \textbf{Value} & \textbf{Purpose} \\
\midrule

\multirow{4}{*}{Warmup Phase}
 & $n_{\text{warmup,min}}$
   & 5
   & Minimum number of function calls executed before checking for stability.
     Ensures early executions do not appear artificially stable. \\
 & $n_{\text{warmup,max}}$
   & 12
   & Maximum number of warmup calls.
     Stops warmup even if stability is not reached. \\
 & $w$
   & 4
   & Number of most recent runtimes used to compute the \gls{cv}.
     Defines the window over which stability is evaluated. \\
 & $\tau_{\text{warmup}}$
   & 0.08
   & Threshold on the \gls{cv}.
     Warmup stops once runtime variability falls below this value. \\
\midrule

\multirow{8}{*}{Runtime calibration}
 & $n_{\text{prime}}$
   & 3
   & Number of untimed invocations executed before the probe phase.
     Removes one-time overhead such as imports and cache initialization. \\

 & $n_{\text{probe,min}}$
   & 3
   & Minimum number of timing samples collected during the probe phase,
     where execution time is estimated. \\

 & $n_{\text{probe,max}}$
   & 10
   & Maximum number of timing samples collected during the probe phase.
     Limits calibration cost for long-running functions. \\

 & $t_{\text{probe,min}}$
   & $1\,\mathrm{s}$
   & Minimum total duration of the probe phase used to estimate the
     representative execution time. \\

 & $t_{\text{window}}$
   & $0.3\,\mathrm{s}$
   & Target duration used to derive the number of invocations per iteration,
     based on the estimated execution time. \\

 & $\alpha$
   & 1.2
   & Safety factor applied during the computation of $n_{\text{inv}}$ to
     compensate for runtime variability and timer noise. \\

 & $n_{\text{inv,min}}$
   & 5
   & Lower bound applied to the computed $n_{\text{inv}}$ before the
     measurement phase. Ensures sufficient work per iteration. \\

 & $n_{\text{inv,max}}$
   & $1\times10^{3}$
   & Upper bound applied to the computed $n_{\text{inv}}$ before the
     measurement phase. Prevents excessive work for very fast functions. \\
\midrule

\multirow{3}{*}{Memory Calibration}
 & $K$
   & 60
   & Target number of memory samples per function call.
     Determines how frequently memory usage is recorded. \\
 & $\delta_{\min}$
   & $2\times10^{-4}\,\mathrm{s}$
   & Minimum time between memory samples.
     Avoids excessive overhead for fast executions. \\
 & $\delta_{\max}$
   & $5\times10^{-2}\,\mathrm{s}$
   & Maximum time between memory samples.
     Ensures sufficient resolution for slower executions. \\
\midrule

\multirow{4}{*}{Adaptive Convergence}
 & $n_{\text{iter,min}}$
   & 5
   & Minimum number of measurement windows collected before checking convergence. \\
 & $n_{\text{iter,max}}$
   & 12
   & Maximum number of measurement windows.
     Stops measurement if convergence is not reached. \\
 & $t_{\text{budget}}$
   & $30\,\mathrm{s}$
   & Maximum total wall-clock time allowed for measurement.
     Terminates measurement when this limit is exceeded. \\
 & $\tau_{\text{CI}}$
   & 0.1
   & Threshold on the relative confidence interval width (\gls{rciw}).
     Measurement stops once results are sufficiently precise. \\
\bottomrule
\end{tabularx}
\end{table}

\subsection{Calibration}
\label{app:calibration}

This subsection describes the calibration procedure used to derive workload-specific measurement parameters. The objective is to obtain runtime and memory measurements that are both statistically stable and representative, while minimizing instrumentation overhead. To this end, calibration determines (i) the number of function invocations per iteration and (ii) the memory sampling interval.

Runtime measurements in microbenchmark settings are inherently noisy due to system-level variability such as scheduling, caching, and background processes. Single executions are therefore insufficient to provide reliable estimates. To address this, each measurement iteration aggregates multiple invocations of the target function, and the median runtime is used as a robust estimator. This reduces the influence of outliers and transient fluctuations. 

We distinguish between \emph{invocations} and \emph{iterations}. An invocation corresponds to a single execution of the target function $f$, whereas an iteration represents one statistical sample used in the evaluation. By grouping multiple invocations into a single iteration, we ensure that each sample has sufficient signal strength while maintaining comparability across iterations.

In contrast, memory measurements obtained via tracemalloc~\cite{PythonTracemalloc} exhibit lower variance but introduce non-negligible overhead due to continuous allocation tracking. Increasing the number of invocations would amplify this overhead. Instead, we calibrate the sampling interval to balance coverage and cost, ensuring that relevant allocation events are captured while limiting instrumentation overhead.

Calibration parameters are inherently system-dependent and therefore not transferable across environments. However, since \bench reports relative performance differences measured under identical conditions, locally calibrated parameters are both necessary and sufficient to ensure comparisons. Table~\ref{tab:measure-config} summarizes the calibration configuration, including timing targets, iteration bounds, and memory sampling parameters.

\begin{algorithm}[!htbp]
\caption{Calibration}
\label{alg:calibration}
\begin{algorithmic}[1]
\Require $f$, calibration parameters (Table~\ref{tab:measure-config})
\Ensure $n_{\text{inv}}$ (runtime measurement invocations per iteration), $\delta$ (memory sampling interval)

\For{$i = 1$ \textbf{to} $n_{\text{prime}}$}
    \State \Call{Restore}{};\enspace $f()$
    \Comment{Warmup}
\EndFor

\State $S \leftarrow [\,]$,\enspace $t_{\text{probe}} \leftarrow 0$
\While{$|S| < n_{\text{probe,min}}$ \textbf{ or }
       $t_{\text{probe}} < t_{\text{probe,min}}$}
    \If{$|S| \geq n_{\text{probe,max}}$} \textbf{break} \EndIf
    \State \Call{Restore}{};\enspace
           $t \leftarrow \mathrm{time}(f())$;\enspace
           $S \leftarrow S \cup \{t\}$;\enspace
           $t_{\text{probe}} \mathrel{+}= t$
\EndWhile
\State $\tilde{t} \leftarrow \mathrm{median}(S)$
       \Comment{Representative runtime estimates}

\State $n_{\text{inv}} \leftarrow \mathrm{clamp}\!\left(
    \left\lceil \dfrac{t_{\text{window}} \cdot \alpha}{\tilde{t}} \right\rceil,\;
    n_{\text{inv,min}},\; n_{\text{inv,max}}
    \right)$
\Comment{Compute invocations per iteration}

\State $\delta \leftarrow \mathrm{clamp}\!\left(
    \dfrac{\tilde{t}}{\max(K, 1)},\;
    \delta_{\min},\; \delta_{\max}
    \right)$
\Comment{Compute memory sampling interval}

\State \Return $n_{\text{inv}},\; \delta$
\end{algorithmic}
\end{algorithm}

The \textsc{Restore} operation reconstructs input data and resets any
modified state before each invocation, ensuring independent and
identical execution conditions. It is not measured.

Before timing begins, $n_{\text{prime}}$ untimed invocations warm up
the Python interpreter and function-level caches, preventing first-call
overhead such as bytecode compilation from affecting measurements.

The probe loop then collects per-invocation timings until a minimum
duration $t_{\text{probe,min}}$ is reached or $n_{\text{probe,max}}$
samples are gathered. From the resulting sample set $S$, two parameters
are derived: the number of invocations per measurement iteration,
\[
    n_{\text{inv}} = \mathrm{clamp}\!\left(
        \left\lceil \frac{t_{\text{window}} \cdot \alpha}{\tilde{t}}
        \right\rceil,\;
        n_{\text{inv,min}},\; n_{\text{inv,max}}
    \right),
\]
and the memory sampling interval,
\[
    \delta = \mathrm{clamp}\!\left(
        \frac{\tilde{t}}{\max(K, 1)},\;
        \delta_{\min},\; \delta_{\max}
    \right).
\]
The median runtime $\tilde{t}$ of the probe sample set is used as a robust estimate of the representative execution time. Based on this estimate, the number of invocations $n_{\text{inv}}$ is chosen such that the total runtime of one measurement iteration approximates the target window duration $t_{\text{window}}$. The safety factor $\alpha$ compensates for measurement overhead and residual runtime variability.

The same representative runtime $\tilde{t}$ is further used to determine the memory sampling interval $\delta$. Using the median instead of the mean ensures robustness to outliers and avoids overly coarse sampling intervals that could miss short-lived allocation patterns.

Calibration is performed once per workload on a baseline implementation. The resulting parameters $n_{\text{inv}}$ and $\delta$ are then reused for all other implementations to ensure consistent and comparable measurements across versions.

\subsection{Warmup}
\label{app:warmup}

The warmup phase repeatedly executes $f$ and evaluates the \gls{cv},
defined as $\mathrm{stdev}(W) / \mathrm{mean}(W)$ over a trailing
window $W$ of the most recent $w$ invocations. As a scale-independent
stability measure, it generalises across workloads with heterogeneous
execution times~\cite{laaber_dynamic_reconfigure}.

Following \cite{georges_rigorous}, warmup targets two properties:
an \emph{initialised} state, where startup effects have subsided, and
an \emph{independent} state, where successive times are approximately
i.i.d. The trailing \gls{cv} serves as an automated proxy for both,
discarding early high-variance observations while detecting local
stability.

Warmup terminates when $\mathrm{CV} < \tau_{\text{warmup}}$, subject
to a minimum of $n_{\text{warmup,min}}$ invocations and a hard cap of
$n_{\text{warmup,max}}$. Algorithm~\ref{alg:warmup} summarises the
procedure.

\begin{algorithm}[H]
\caption{Warmup}
\label{alg:warmup}
\begin{algorithmic}[1]
\Require $f$, window $w$, threshold $\tau_{\text{warmup}}$,
         bounds $n_{\text{warmup,min}},\; n_{\text{warmup,max}}$
\Ensure convergence before measurement

\State $T \leftarrow [\,]$
\For{$i \gets 1$ \textbf{to} $n_{\text{warmup,max}}$}
    \State \Call{Restore}{};\enspace
           $T \leftarrow T \cup \{\mathrm{runtime}(f())\}$
    \If{$|T| \geq \max(n_{\text{warmup,min}},\; w)$}
        \State $\mathrm{CV} \leftarrow
               \mathrm{stdev}(T_{[-w:]}) / \mathrm{mean}(T_{[-w:]})$
               \Comment{trailing window over last $w$ samples}
        \If{$\mathrm{CV} < \tau_{\text{warmup}}$} \textbf{break} \EndIf
    \EndIf
\EndFor
\State \Return $\mathrm{CV} < \tau_{\text{warmup}}$
\end{algorithmic}
\end{algorithm}

\subsection{Measurement}
\label{app:measurement-phases}

 After warmup convergence, each workload is evaluated over multiple iterations, each consisting of two separated phases to avoid instrumentation interference. Measurement is performed per workload in an isolated container using repeated measurement iterations. Each iteration consists of a runtime phase and a memory phase.

\emph{Runtime phase.}
 Within each iteration, the target function is executed multiple times, as determined during calibration, and the median of these executions yields a single, stable runtime measurement. These per-iteration measurements are then used for convergence analysis.

\emph{Memory phase.}
 A single invocation is executed under \texttt{tracemalloc}~\cite{PythonTracemalloc}
 with a background sampler, where the sampling rate is calibrated based on the function runtime to ensure sufficient temporal resolution. From the sampled values, we compute the \gls{twmu}, while peak memory usage above baseline is obtained directly with \texttt{tracemalloc}.

\paragraph{Adaptive Convergence}

Iterations are collected until one of the following stopping conditions is met:
(1) the iteration count reaches a predefined maximum,
(2) the total measurement time exceeds a predefined limit after at least a minimum number of samples has been collected, or
(3) a minimum number of samples has been collected and all monitored metrics satisfy a target \gls{rciw}.

For a sample vector $x = (x_1,\dots,x_n)$, the \gls{rciw} is defined as
\[
\mathrm{RCIW}(x)
=
\frac{\mathrm{CI}_{1-\alpha}(x)^{\text{upper}} - \mathrm{CI}_{1-\alpha}(x)^{\text{lower}}}
{\mathrm{mean}(x)}.
\]

The $(1-\alpha)$ confidence interval of the mean is given by
\[
\mathrm{CI}_{1-\alpha}(x)
=
\left[
\mathrm{mean}(x) - q_{\alpha/2} \cdot \frac{\mathrm{stdev}(x)}{\sqrt{n}},
\;
\mathrm{mean}(x) + q_{\alpha/2} \cdot \frac{\mathrm{stdev}(x)}{\sqrt{n}}
\right],
\]
where $q_{\alpha/2}$ denotes the critical value of the standard normal distribution corresponding to the chosen confidence level.

Let $r = (r_1,\dots,r_n)$ denote the per-iteration runtime samples, where each $r_i$ is computed as the median runtime over multiple function invocations within iteration $i$. 
The number of function invocations per iteration is determined during a prior calibration phase to reduce measurement noise and ensure stable estimates for \gls{rciw}. Similarly, let $m^{\mathrm{twmu}} = (m^{\mathrm{twmu}}_1,\dots,m^{\mathrm{twmu}}_n)$ denote the time-weighted memory usage samples, and $m^{\mathrm{peak}} = (m^{\mathrm{peak}}_1,\dots,m^{\mathrm{peak}}_n)$ the peak memory increase samples.

Adaptive sampling terminates once
\[
\mathrm{RCIW}(r) \le \tau_{\text{CI}}, \qquad
\mathrm{RCIW}(m^{\mathrm{twmu}}) \le \tau_{\text{CI}}, \qquad
\mathrm{RCIW}(m^{\mathrm{peak}}) \le \tau_{\text{CI}}.
\]
where $\tau_{\mathrm{CI}}$ denotes the convergence threshold.

Condition~(3) serves as the primary stopping criterion, ensuring that all reported metrics satisfy a predefined relative error bound at the chosen confidence level. The full measurement procedure is shown in Algorithm~\ref{alg:measurement}.

\begin{algorithm}[t]
\caption{Adaptive Measurement}
\label{alg:measurement}
\begin{algorithmic}[1]
\Require target function $f$,
runtime invocations $n_{\text{inv}}$,
sampling interval $\delta$,
convergence threshold $\tau_{\text{CI}}$,
bounds $n_{\text{iter,min}}, n_{\text{iter,max}}$,
time budget $t_{\text{budget}}$
\Ensure sample set $\mathcal{I}$

\State \Call{Warmup}{$f$}
\State $\mathcal{I} \gets [\,]$
\State $t_{\text{start}} \gets \mathrm{now}()$

\Loop
  \State $R \gets [\,]$
  \State \Call{DisableGC}{}
  \For{$j = 1$ \textbf{to} $n_{\text{inv}}$}
    \State \Call{Restore}{}
    \State append $R \gets \mathrm{runtime}(f())$
  \EndFor
  \State \Call{EnableGC}{}

  \State $r \gets \mathrm{median}(R)$

  \State \Call{Restore}{}
  \State \Call{StartMemoryTracer}{}
  \State \Call{StartBackgroundSampler}{$\delta$}
  \State $f()$
  \State \Call{StopBackgroundSampler}{}
  \State \Call{StopMemoryTracer}{}
  \State $m_{\text{peak}}, m_{\text{twmu}} \gets \Call{ComputeMemoryMetrics}{\cdot}$

  \State append $\mathcal{I} \gets (r, m_{\text{peak}}, m_{\text{twmu}})$
  \State $n \gets |\mathcal{I}|$
  \State $t_{\text{elapsed}} \gets \mathrm{now}() - t_{\text{start}}$

  \If{$n \ge n_{\text{iter,max}}$}
    \State \textbf{break}
  \EndIf

  \If{$n \ge n_{\text{iter,min}}$}
    \State $R \gets \{r_i\}$
    \State $M_{\text{twmu}} \gets \{m^{\text{twmu}}_i\}$
    \State $M_{\text{peak}} \gets \{m^{\text{peak}}_i\}$

    \If{$\mathrm{RCIW}(R) \le \tau_{\text{CI}}$ \textbf{and}
        $\mathrm{RCIW}(M_{\text{twmu}}) \le \tau_{\text{CI}}$ \textbf{and}
        $\mathrm{RCIW}(M_{\text{peak}}) \le \tau_{\text{CI}}$}
        \State \textbf{break}
    \EndIf

    \If{$t_{\text{elapsed}} \ge t_{\text{budget}}$}
        \State \textbf{break}
    \EndIf
  \EndIf
\EndLoop

\State \Return $\mathcal{I}$
\end{algorithmic}
\end{algorithm}

\subsection{Validation Aggregation}
\label{app:validation_aggregation}

We aggregate validation results in two stages: first within each \gls{pr}, and then across \glspl{pr}. This separation reflects the distinction between workloads belonging to the same optimization context and independent optimization cases originating from different \glspl{pr}.

Within a \gls{pr}, each workload configuration is evaluated independently. Every workload consists of multiple container runs, and each run produces a sequence of profiling samples for each performance metric. To reduce the influence of transient measurement artifacts, we apply \gls{mad}-based outlier trimming. For a set of observations \(x_1,\dots,x_n\), the \gls{mad} is defined as:

\[
\mathrm{MAD} =
\mathrm{median}
\left(
|x_i - \mathrm{median}(x_1,\dots,x_n)|
\right)
\]

An observation \(x_i\) is retained if its deviation from the median satisfies:

\[
|x_i - \mathrm{median}(x)| \leq k \cdot c \cdot \mathrm{MAD}
\]

where \(k\) is a configurable sensitivity parameter and \(c\) is the consistency factor used to align the estimator with the standard deviation under normality assumptions. 

The \gls{mad} trimming procedure is applied at two levels: within containers and between containers. Profiling samples inside each container run are first filtered to reduce transient effects, after  across container-level summaries to reduce anomalous runs caused by external system variability. To ensure reliability, a workload is considered stable only when the \gls{rciw} of both the baseline and \gls{llm}-optimized measurements remains below the configured system-noise threshold. Stability is evaluated at two granularities: within-run variability and between-run variability.

Stable workloads are subsequently filtered using a \gls{snr}:

\[
\mathrm{SNR} =
\frac{|\Delta\%|}
{\max \left(
\mathrm{RCIW}_{\mathrm{within\text{-}baseline}},
\mathrm{RCIW}_{\mathrm{within\text{-}llm}},
\mathrm{RCIW}_{\mathrm{between\text{-}baseline}},
\mathrm{RCIW}_{\mathrm{between\text{-}llm}}
\right)}
\]

where \(\Delta\%\) denotes the relative performance change. A workload is retained only if the observed effect sufficiently exceeds the estimated measurement noise. Following prior recommendations, we use an \gls{snr} threshold of \(2\), corresponding to a conservative detectability boundary between signal and noise~\cite{snr2}.

The remaining workloads are classified as either improvements, regressions and conflicting. Since \glspl{if} represent multiplicative, rate-like changes, workload-level \glspl{if} within a \gls{pr} are aggregated using the harmonic mean, following recommendations for summarizing normalized performance ratios~\cite{harmonic}. Given workload ratios \(r_1,\dots,r_n\), the \gls{pr}-level summary is computed as:

\[
H_{\mathrm{PR}} =
\frac{n}
{\sum_{i=1}^{n} \frac{1}{r_i}}
\]

Only workloads that satisfy the stability and \gls{snr} criteria contribute to this aggregation.

Each metric for a given \gls{pr} is then classified according to the following hierarchy:

\begin{itemize}
    \item \textbf{Not measurable}: The proportion of stable workloads falls below the minimum quality requirement.
    
    \item \textbf{No signal}: No stable workload passes the \gls{snr} criterion.
    
    \item \textbf{Improved}: Improvements dominate and no regressions are observed.
    
    \item \textbf{Regressed}: Regressions dominate and no improvements are observed.
    
    \item \textbf{Conflicting}: Both improvements and regressions exceed the configured conflict tolerance.
\end{itemize}

After computing \gls{pr}-level summaries, results are aggregated across \glspl{pr}. Each \gls{pr} classified as improved or regressed is treated as one independent optimization case. The final cross-\gls{pr} estimate is computed as the arithmetic mean of the corresponding \gls{pr}-level harmonic means:

\[
S =
\frac{1}{m}
\sum_{j=1}^{m}
H_{\mathrm{PR},j}
\]

where \(m\) denotes the number of eligible \glspl{pr}. This two-level aggregation strategy reflects the different semantic roles of workloads and \glspl{pr}. The harmonic mean captures the joint effect of related workloads within a single optimization context, while the arithmetic mean summarizes independent optimization cases across \glspl{pr}. By excluding unstable measurements prior to aggregation, the framework ensures that both workload-level and cross-\gls{pr} estimates remain robust against measurement noise and transient execution artifacts.

\subsection{System Configuration and Execution Environment}
\label{app:docker}
%DONE
To ensure reproducible performance evaluation, \bench relies on
pre-built Docker images for both baseline and optimized versions of
each scenario. These images are constructed once and stored with
immutable identifiers (tags and digests), guaranteeing that all
experiments are executed in identical environments.

\paragraph{Docker setup}
Each benchmark is executed within a resource-constrained Docker
container with fixed resource and runtime settings:

\begin{itemize}
    \item \textbf{CPU pinning:} Containers are restricted to a fixed CPU
    core (\texttt{cpuset=2}) and memory node
    (\texttt{cpuset\_mems=0})

    \item \textbf{Memory limit:} A fixed memory budget
    (\texttt{30GB}) is enforced across runs

    \item \textbf{Thread control:}
    Numerical backends are restricted to single-threaded execution using
    \texttt{OMP\_NUM\_THREADS=1},
    \texttt{OPENBLAS\_NUM\_THREADS=1},
    \texttt{MKL\_NUM\_THREADS=1},
    \texttt{NUMEXPR\_MAX\_THREADS=1},
    \texttt{VECLIB\_MAXIMUM\_THREADS=1},
    and \texttt{BLIS\_NUM\_THREADS=1}, ensuring consistent execution
    behavior across runs
\end{itemize}

\paragraph{Experimental Setup}

All experiments were conducted on a MacBook Pro equipped with 
an Apple M4 Max chip (14-core CPU with 10 performance and 4 efficiency cores) 
and 36\,GB unified memory. Patch application and evaluation were executed in 
isolated Docker containers running on the same host.

\paragraph{Image Versioning}
Rather than rebuilding environments during benchmarking, \bench
uses pre-built Docker images for both baseline and optimized versions.
These images are stored in a registry and referenced via immutable
digests.

\subsection{Additional Results}
\label{additional_results}
Under this subsection, we present additional results that did not fit in the main section. Table~\ref{tab:results-correctness} reports patch generation and correctness success rates under both Oracle and BM25 retrieval settings.

\begin{table}[ht]
\centering
\small
\caption{Patch generation and correctness pass rate under Oracle and BM25
retrieval. Patch denotes the successful patch application rate (\%);
Test denotes the percentage of all evaluated samples that pass all relevant tests (\%).}
\label{tab:results-correctness}
\begin{tabular}{lcccc}
\toprule
& \multicolumn{2}{c}{\textbf{Oracle}}
& \multicolumn{2}{c}{\textbf{BM25}} \\
\cmidrule(lr){2-3} \cmidrule(lr){4-5}
\textbf{Model}
& \textbf{Patch@\cmark} & \textbf{Test@\cmark}
& \textbf{Patch@\cmark} & \textbf{Test@\cmark} \\
\midrule
GPT-5.2           & 95.1\% &79.4\% & 86.3\% & 68.6\% \\
Claude Sonnet 4.6 & 90.2\% & 72.5\% & 70.6\% & 55.9\%  \\
Gemini 3.1 Flash  & 93.1\% & 50.0\% & 78.4\% & 57.8\%  \\
Kimi K2.5         & 30.4\% & 18.6\% & 17.6\% & 11.8\%  \\
GLM-5.1           & 97.1\% & 55.9\% & 63.7\% & 39.2\% \\
MiniMax M2.7      & 45.1\% & 23.5\% & 21.6\% & 8.8\%  \\
\bottomrule
\end{tabular}
\end{table}

\subsubsection{Claude Sonnet 4.6 Optimization Analysis for PR~\#7578}
\label{app:pr7578}

\begin{diffbox}{Claude Sonnet 4.6 patch with BM25 retrieval for PR~\#7578}
diff --git a/xarray/core/rolling.py b/xarray/core/rolling.py
index d49cb6e1..c0749e2b 100644
--- a/xarray/core/rolling.py
+++ b/xarray/core/rolling.py
@@ -770,12 +770,11 @@ class DatasetRolling(Rolling["Dataset"]):
             dims = [d for d in self.dim if d in da.dims]
             if dims:
                 wi = {d: window_dims[i] for i, d in enumerate(self.dim) if d in da.dims}
-                st = {d: strides[i] for i, d in enumerate(self.dim) if d in da.dims}
 
                 dataset[key] = self.rollings[key].construct(
                     window_dim=wi,
                     fill_value=fill_value,
-                    stride=st,
+                    stride=1,
                     keep_attrs=keep_attrs,
                 )
             else:
@@ -787,9 +786,14 @@ class DatasetRolling(Rolling["Dataset"]):
 
         attrs = self.obj.attrs if keep_attrs else {}
 
-        return Dataset(dataset, coords=self.obj.coords, attrs=attrs).isel(
-            {d: slice(None, None, s) for d, s in zip(self.dim, strides)}
-        )
+        result = Dataset(dataset, coords=self.obj.coords, attrs=attrs)
+
+        if any(s != 1 for s in strides):
+            result = result.isel(
+                {d: slice(None, None, s) for d, s in zip(self.dim, strides)}
+            )
+
+        return result
\end{diffbox}

\begin{diffbox}{Claude Sonnet 4.6 patch with Oracle retrieval for PR~\#7578}
diff --git a/xarray/core/rolling.py b/xarray/core/rolling.py
index d49cb6e1..f1393c57 100644
--- a/xarray/core/rolling.py
+++ b/xarray/core/rolling.py
@@ -787,9 +787,21 @@ class DatasetRolling(Rolling["Dataset"]):
 
         attrs = self.obj.attrs if keep_attrs else {}
 
-        return Dataset(dataset, coords=self.obj.coords, attrs=attrs).isel(
-            {d: slice(None, None, s) for d, s in zip(self.dim, strides)}
-        )
+        # Apply stride to coordinates along rolling dimensions.
+        # Data variables already have stride applied inside each
+        # DataArrayRolling.construct call, so we must NOT apply isel again
+        # on the whole Dataset (that would double-stride data variables).
+        stride_indexer = {d: slice(None, None, s) for d, s in zip(self.dim, strides)}
+        new_coords = {
+            name: coord.isel(
+                {d: stride_indexer[d] for d in stride_indexer if d in coord.dims}
+            )
+            if any(d in coord.dims for d in stride_indexer)
+            else coord
+            for name, coord in self.obj.coords.items()
+        }
+
+        return Dataset(dataset, coords=new_coords, attrs=attrs)
\end{diffbox}

\begin{figure}[H] 
    \centering
    \includegraphics[width=0.9\linewidth]{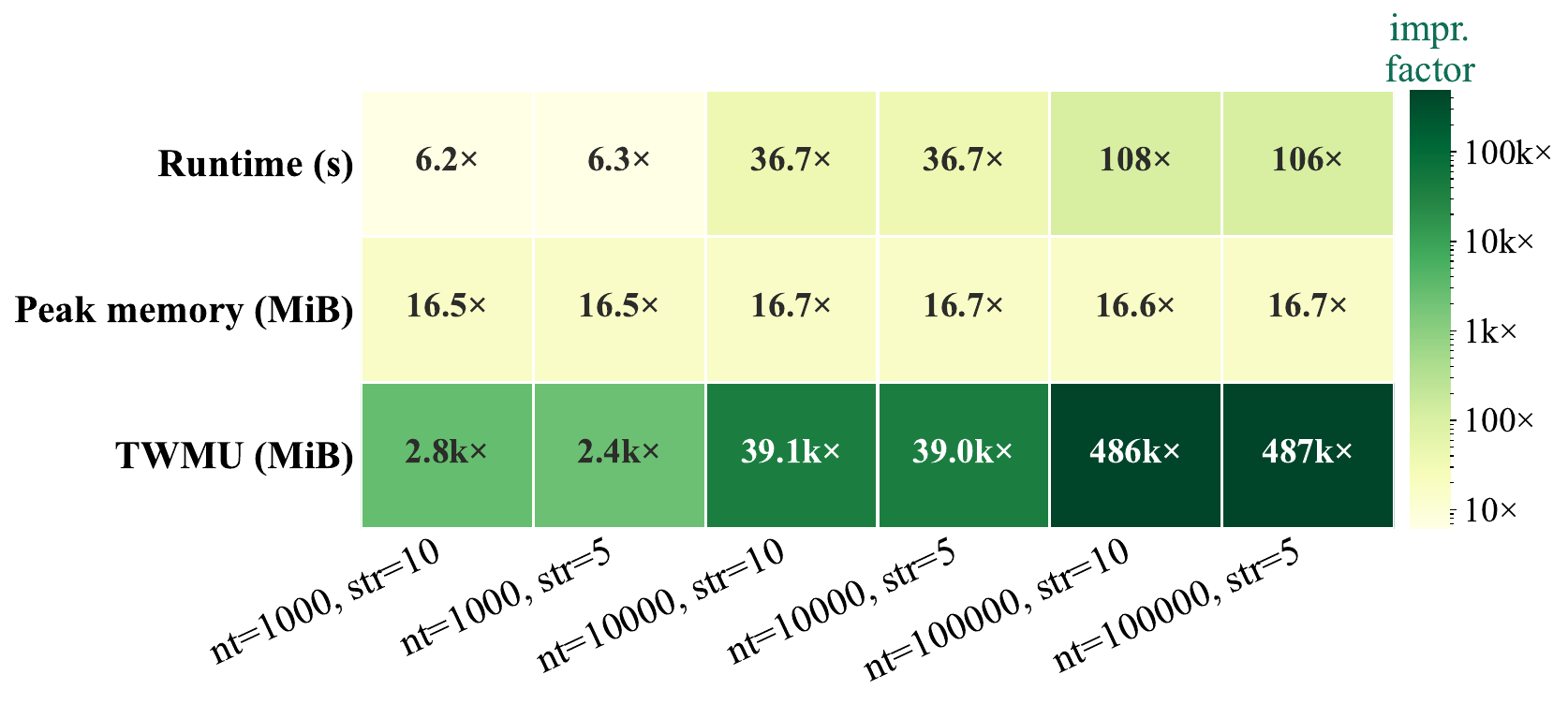}
    \caption{
Impact of input configuration on optimization effectiveness for PR~\#7578 under BM25 retrieval.
Per-workload \gls{if} varies with rolling size (\texttt{nt}) and stride
(\texttt{s}), highlighting input-dependent performance behavior.}
    \label{fig:llm_anaysis_claude_bm25}
\end{figure}

Figure~\ref{fig:llm_anaysis_claude_bm25} presents the workload-level optimization behavior of the BM25-retrieved context with Claude Sonnet 4.6 for PR~\#7578, while the complementary Oracle-retrieved result is shown in Figure~\ref{fig:llm_anaysis_claude_oracle}. The workload configurations and performance tests used for this \gls{pr} in this analysis are described in Section~\ref{par:xarray_7578}.

Both patches target the same inefficiency in the rolling-window construction pipeline caused by repeated stride applications during dataset construction. The BM25-retrieved patch simplifies the execution flow by enforcing \texttt{stride=1} during rolling construction and applying slicing only once afterward. In contrast, the Oracle-retrieved patch applies stride handling more selectively by operating only on coordinates and avoiding double-striding of data variables.

As shown in Figure~\ref{fig:llm_anaysis_claude_bm25}, the BM25-retrieved patch achieves a stronger optimization effect on \gls{twmu} for workload \texttt{nt=10000}, with runtime improvements exceeding \(100\times\) and \gls{twmu} reductions approaching \(490\times\). The Oracle-retrieved patch exhibits similar optimization trends in Figure~\ref{fig:llm_anaysis_claude_oracle}, suggesting that the main performance gains originate from eliminating repeated stride operations rather than from coordinate-handling logic itself. The \texttt{nt=10000} workload thus serves as the decisive differentiator in this evaluation.
This comparison also illustrates how optimization behavior can vary considerably across different execution configurations with different solutions.

\end{document}